\documentclass[a4paper,USenglish,cleveref,autoref,thm-restate]{lipics-v2021}

\pdfoutput=1 
\hideLIPIcs  

\title{Sampling Unlabeled Chordal Graphs in Expected Polynomial Time}

\author{\'Ursula H\'ebert-Johnson}{University of California, Santa Barbara, USA}{ursula@ucsb.edu}{}{Supported by NSF grant CCF-2147094.}

\author{Daniel Lokshtanov}{University of California, Santa Barbara, USA}{daniello@ucsb.edu}{}{}

\authorrunning{\'U. H\'ebert-Johnson and D. Lokshtanov}

\Copyright{\'Ursula H\'ebert-Johnson and Daniel Lokshtanov}

\ccsdesc[500]{Theory of computation~Generating random combinatorial structures}
\ccsdesc[500]{Theory of computation~Graph algorithms analysis}

\keywords{Chordal graphs, graph sampling, graph counting, unlabeled graphs}

\relatedversion{}
\relatedversiondetails{Full Version}{https://arxiv.org/abs/2501.05024}

\acknowledgements{We would like to thank Eric Vigoda for discussions that led to a concise proof of an important step in the running time analysis.}

\nolinenumbers

\usepackage{mathtools}
\usepackage{algorithm}
\usepackage{algpseudocode}

\newcommand{\N}{\mathbb{N}}
\newcommand{\E}{\mathbf{E}}

\newcommand{\GammaC}{\Gamma_{\text{chord}}}
\newcommand{\OmegaC}{\Omega_{\text{chord}}}

\newcommand{\g}[8]{g\!
\begin{pmatrix}
#1\, & #2 & #3 & #4 \\
#5\, & #6 & #7 & #8
\end{pmatrix}}
\newcommand{\G}[8]{G\!
\begin{pmatrix}
#1\, & #2 & #3 & #4 \\
#5\, & #6 & #7 & #8
\end{pmatrix}}

\newcommand{\gTilde}[8]{\tilde g\!
\begin{pmatrix}
#1\, & #2 & #3 & #4 \\
#5\, & #6 & #7 & #8
\end{pmatrix}}
\newcommand{\GTilde}[8]{\widetilde G\!
\begin{pmatrix}
#1\, & #2 & #3 & #4 \\
#5\, & #6 & #7 & #8
\end{pmatrix}}

\newcommand{\gTildeP}[8]{\tilde g_p\!
\begin{pmatrix}
#1\, & #2 & #3 & #4 \\
#5\, & #6 & #7 & #8
\end{pmatrix}}
\newcommand{\GTildeP}[8]{\widetilde G_p\!
\begin{pmatrix}
#1\, & #2 & #3 & #4 \\
#5\, & #6 & #7 & #8
\end{pmatrix}}

\newcommand{\gTildeOne}[6]{\tilde g_1\!
\begin{pmatrix}
#1\, & #2 & #3 \\
#4\, & #5 & #6
\end{pmatrix}}
\newcommand{\GTildeOne}[6]{\widetilde G_1\!
\begin{pmatrix}
#1\, & #2 & #3 \\
#4\, & #5 & #6
\end{pmatrix}}

\newcommand{\gTildeTwo}[6]{\tilde g_{\geq 2}\!
\begin{pmatrix}
#1\, & #2 & #3 \\
#4\, & #5 & #6
\end{pmatrix}}
\newcommand{\GTildeTwo}[6]{\widetilde G_{\geq 2}\!
\begin{pmatrix}
#1\, & #2 & #3 \\
#4\, & #5 & #6
\end{pmatrix}}

\newcommand{\f}[8]{f\!
\begin{pmatrix}
#1\, & #2 & #3 & #4 \\
#5\, & #6 & #7 & #8
\end{pmatrix}}
\newcommand{\F}[8]{F\!
\begin{pmatrix}
#1\, & #2 & #3 & #4 \\
#5\, & #6 & #7 & #8
\end{pmatrix}}

\newcommand{\fTilde}[8]{\tilde f\!
\begin{pmatrix}
#1\, & #2 & #3 & #4 \\
#5\, & #6 & #7 & #8
\end{pmatrix}}
\newcommand{\FTilde}[8]{\widetilde F\!
\begin{pmatrix}
#1\, & #2 & #3 & #4 \\
#5\, & #6 & #7 & #8
\end{pmatrix}}

\newcommand{\fTildeP}[8]{\tilde f_p\!
\begin{pmatrix}
#1\, & #2 & #3 & #4 \\
#5\, & #6 & #7 & #8
\end{pmatrix}}
\newcommand{\FTildeP}[8]{\widetilde F_p\!
\begin{pmatrix}
#1\, & #2 & #3 & #4 \\
#5\, & #6 & #7 & #8
\end{pmatrix}}

\newcommand{\neworrenewcommand}[1]{\providecommand{#1}{}\renewcommand{#1}}

\newcommand{\fTildePWithZ}[9]{
    \neworrenewcommand{\fTildePWithZHelper}[1]{
        \tilde f_p\!
        \begin{pmatrix}
        #1\, & #2 & #3 & #4 & #5 \\
        #6\, & #7 & #8 & #9 & ##1
        \end{pmatrix}
    }
    \fTildePWithZHelper
}
\newcommand{\FTildePWithZ}[9]{
    \neworrenewcommand{\FTildePWithZHelper}[1]{
        \widetilde F_p\!
        \begin{pmatrix}
        #1\, & #2 & #3 & #4 & #5 \\
        #6\, & #7 & #8 & #9 & ##1
        \end{pmatrix}
    }
    \FTildePWithZHelper
}

\mathchardef\standardl=\mathcode`l
\mathcode`l=\ell
\newcommand{\deactivatel}{\mathcode`l=\standardl}
\makeatletter
\edef\operator@font{\operator@font\noexpand\deactivatel}
\makeatother

\EventEditors{John Q. Open and Joan R. Access}
\EventNoEds{2}
\EventLongTitle{42nd Conference on Very Important Topics (CVIT 2016)}
\EventShortTitle{CVIT 2016}
\EventAcronym{CVIT}
\EventYear{2016}
\EventDate{December 24--27, 2016}
\EventLocation{Little Whinging, United Kingdom}
\EventLogo{}
\SeriesVolume{42}
\ArticleNo{23}

\begin{document}

\maketitle

\begin{abstract}
We design an algorithm that generates an $n$-vertex unlabeled chordal graph uniformly at random in expected polynomial time. Along the way, we develop the following two results: (1) an $\mathsf{FPT}$ algorithm for counting and sampling labeled chordal graphs with a given automorphism $\pi$, parameterized by the number of moved points of $\pi$, and (2) a proof that the probability that a random $n$-vertex labeled chordal graph has a given automorphism $\pi\in S_n$ is at most $1/2^{c\max\{\mu^2,n\}}$, where $\mu$ is the number of moved points of $\pi$ and $c$ is a constant. Our algorithm for sampling unlabeled chordal graphs calls the aforementioned $\mathsf{FPT}$ algorithm as a black box with potentially large values of the parameter $\mu$, but the probability of calling this algorithm with a large value of $\mu$ is exponentially small.
\end{abstract}

\section{Introduction}

A graph is \emph{chordal} if it has no induced cycles of length at least 4. The term was coined by Gavril in 1972~\cite{gavril1972algorithms}, more than fifty years ago, but the notion of chordal graphs in fact goes as far back as 1958~\cite{hajnal1958auflosung}. In the early papers, chordal graphs were referred to by other names, such as \emph{triangulated} graphs. By now, many structural results have been proved about chordal graphs, and there are many algorithms that take a chordal graph as input. Thus it would be useful to have an efficient algorithm for generating random chordal graphs, both for the practical purpose of software testing, as well as the more mathematical purpose of testing conjectures.

In~\cite{hebertjohnson2023counting}, H\'ebert-Johnson et al.\ designed an algorithm that generates $n$-vertex labeled chordal graphs uniformly at random. This algorithm runs in polynomial time, using at most $O(n^7)$ arithmetic operations for the first sample and $O(n^4)$ arithmetic operations for each subsequent sample. However, when discussing the performance of an algorithm that is being tested, the correct output typically does not depend on the labeling of the vertices. If we use a labeled-graph sampling algorithm to generate random test cases, then asymmetric graphs will be given too much weight/probability compared to those that happen to have many automorphisms. This naturally leads to the question of efficiently generating \emph{unlabeled} chordal graphs uniformly at random. In this paper, we present an algorithm that solves this problem and runs in expected polynomial time.

\begin{restatable}{theorem}{unlabeled}
There is a randomized algorithm that given $n\in\N$, generates a graph uniformly at random from the set of all unlabeled chordal graphs on $n$ vertices. This algorithm uses $O(n^7)$ arithmetic operations in expectation.
\end{restatable}

It is worth mentioning the difference between the running times for labeled vs.\ unlabeled chordal graph sampling. The sampling algorithm of H\'ebert-Johnson et al.\ generates a random $n$-vertex labeled chordal graph in polynomial time, even in the worst case. However, obtaining a worst-case polynomial-time algorithm for sampling \emph{unlabeled} chordal graphs --- or an expected-polynomial-time counting algorithm for such graphs --- appears to be very difficult since these questions remain open even for general graphs. This is relevant because the class of chordal graphs is known to be $\mathsf{GI}$-complete. While there exist classes of graphs (which we discuss below) for which efficient unlabeled sampling algorithms are known, we are not aware of any $\mathsf{GI}$-complete graph class with such an algorithm.

Our algorithm for sampling unlabeled chordal graphs builds upon an algorithm of Wormald that generates $n$-vertex unlabeled graphs uniformly at random in expected time $O(n^2)$~\cite{wormald1987generating}. This in turn builds upon a related algorithm by Dixon and Wilf~\cite{dixon1983random} that solves the same problem but assumes that the exact number of $n$-vertex unlabeled graphs has already been computed. In~\cite{wormald1987generating}, the algorithm of Wormald follows a somewhat similar structure but removes that assumption. As mentioned above, the question of computing the exact number of $n$-vertex unlabeled graphs in expected polynomial time remains open to this day.

It often happens that we wish to sample from a particular graph class (e.g., chordal graphs). For unlabeled trees, there is a uniform sampling algorithm that runs in polynomial time \cite{wilf1981uniform}. One can also count the exact number of unlabeled trees on $n$ vertices in polynomial time~\cite[A000055]{oeis}. On the topic of counting, an algorithm for counting unlabeled $k$-trees is presented in~\cite{gainer2014counting}, but the running time is not stated. An expected-polynomial-time algorithm for uniform sampling of $2$-connected unlabeled planar graphs was presented by Bodirsky et al.\ in 2005~\cite{bodirsky2005sampling}, followed by the same result for connected unlabeled cubic planar graphs in 2008~\cite{bodirsky2008generating}. For the class of connected unlabeled bipartite permutation graphs, a uniform sampling algorithm was designed by Saitoh et al.\ that runs in $O(n)$ time~\cite{saitoh2012random}.

Although extensive research has been done on the topic of labeled graph sampling \cite{bodirsky2007generating,fusy2009uniform,GW-regular,GW-power}, to the best of our knowledge, the literature on sampling \emph{unlabeled} graphs from a given graph class is relatively sparse. As is the case for chordal graphs, the corresponding labeled sampling problem tends to be solved first for a given graph class, and then perhaps one can address the problem of efficiently sampling unlabeled graphs from the same graph class.





\subsection{Methods}

The sampling algorithm of Wormald~\cite{wormald1987generating} is based on the fact that unlabeled graphs correspond to orbits of the following group action: the symmetric group $S_n$ acts on the set of labeled graphs by permuting the vertex labels. This correspondence follows from the Frobenius-Burnside lemma. Therefore, to sample a random unlabeled graph, it is enough to sample a random orbit of this group action.

To make this approach work for chordal graphs, we need two ingredients: (1) an algorithm for counting and sampling labeled chordal graphs with a given automorphism $\pi$, and (2) a proof that the probability that a random $n$-vertex labeled chordal graph has a given automorphism $\pi\in S_n$ is at most $1/2^{c\max\{\mu^2,n\}}$, where $\mu$ is the number of moved points of $\pi$ and $c$ is a constant.

For (1), we design an $\mathsf{FPT}$ (fixed-parameter tractable) counting algorithm that is parameterized by $\mu$, the number of moved points of $\pi$. This algorithm uses $O(2^{7\mu}n^9)$ arithmetic operations. Using the standard sampling-to-counting reduction of~\cite{JVV}, we also obtain a corresponding sampling algorithm with the same running time. Our main algorithm (for sampling \emph{unlabeled} chordal graphs) calls each of these $\mathsf{FPT}$ algorithms as a black box with potentially large values of the parameter $\mu$. Nevertheless, using the bound from (2), we are able to show that the probability of using a large value of $\mu$ is exponentially small, so the expected running time is not significantly affected.

To design the counting algorithm for (1), we rewrite each of the recurrences from the algorithm of H\'ebert-Johnson et al., now carrying around information about the automorphism $\pi$ and its moved points. The original algorithm is a dynamic-programming algorithm in which there is a constant number of types of vertices ($X$, $L$, etc.) in each graph that we wish to count. In our updated version, we now keep track of the type of each vertex that is moved by $\pi$.

To prove the bound for (2), we distinguish between the cases when $\mu$ is small and $\mu$ is large ($\mu$ is either less than or greater than $\frac{n}{d\log n}$, where $d$ is a constant). When $\mu$ is small, we use the fact that almost every chordal graph is a split graph~\cite{bender1985almost}, and we strengthen this by showing that in fact, almost every chordal graph is a balanced split graph. For the case of balanced split graphs, the argument is easy and is similar to the proof of the bound for general graphs~\cite{oberschelp1967kombinatorische}. When $\mu$ is large, the argument is more complicated. We observe that the vast majority of $n$-vertex labeled chordal graphs have maximum clique size close to $n/2$. Along the way, we also use the fact that for every chordal graph $G$, there exists a PEO (perfect elimination ordering) of $G$ such that some maximum clique appears at the tail end of that PEO.



\section{Preliminaries}


Let $\N$ be the set of natural numbers, not including 0. For $n\in\N$, we use the notation $[n]\coloneqq\{1,2,\ldots,n\}$. For a graph $G$ and vertex subsets $S,T\subseteq V(G)$, we say $S$ \emph{sees all of} $T$ if $T\subseteq N(S)$.

\begin{definition}
Let $A = \{a_1,\ldots,a_r\}$ and $B = \{b_1,\ldots,b_r\}$ be finite subsets of $\N$ such that $|A| = |B|$, where the elements $a_i$ and $b_i$ are listed in increasing order. We define $\phi(A,B)\colon A\to B$ as the bijection that maps $a_i$ to $b_i$ for all $i\in[r]$.
\end{definition}

\subsection{Permutations and labeled graphs}

For $n\in\N$, let $S_n$ denote the group of all permutations of $[n]$. For a permutation $\pi\in S_n$, we define $M_{\pi}\coloneqq\{i\in[n]:\pi(i)\ne i\}$ to be the set of points moved by $\pi$. For $n\in\N$, $[n]_0\coloneqq\{0,2,3,\ldots,n\}$ denotes the set of all possible values of $|M_{\pi}|$ for $\pi\in S_n$.

Suppose $\pi\in S_n$, $C\subseteq [n]$. We write $\pi(C)\coloneqq\{\pi(i):i\in C\}$ to denote the image of $C$ under $\pi$. We say $C$ is \emph{invariant} under $\pi$ if $\pi(i)\in C$ for all $i\in C$. For a set $C$ that is invariant under $\pi$, we write $\pi|_C$ to denote the permutation $\pi$ restricted to the domain $C$.

A \emph{labeled graph} is a pair $G = (V,E)$, where the vertex set $V$ is a finite subset of $\N$ and the edge set $E$ is a set of two-element subsets of $V$. For a permutation $\pi\in S_n$ and a labeled graph $G$ such that $V(G)\subseteq[n]$ is invariant under $\pi$, we say $\pi|_{V(G)}$ is an \emph{automorphism} of $G$ if for all $u,v\in V(G)$, $u$ and $v$ are adjacent if and only if $\pi(u)$ and $\pi(v)$ are adjacent.

\subsection{Chordal graphs and related notions}

A vertex $v$ in a graph $G$ is \emph{simplicial} if its neighborhood $N(v)$ is a clique. A \emph{perfect elimination ordering} (PEO) of a graph $G$ is an ordering $v_1,\ldots,v_n$ of the vertices of $G$ such that for all $i\in[n]$, $v_i$ is simplicial in the subgraph induced by the vertices $v_i,\ldots,v_n$. A graph is chordal if and only if it has a perfect elimination ordering~\cite{blair1993introduction}. The following two lemmas are well-known facts about chordal graphs. The proof of the first can be found in \cite{blair1993introduction}.

\begin{lemma}
\label{lemma:simplicial}
Every chordal graph $G$ contains a simplicial vertex. If $G$ is not a complete graph, then $G$ contains two non-adjacent simplicial vertices.
\end{lemma}

\begin{lemma}
\label{lemma:maximal_cliques}
A chordal graph $G$ on $n$ vertices has at most $n$ maximal cliques.
\end{lemma}

\begin{proof}
Let $C$ be a maximal clique in $G$, and let $v_i$ be the leftmost vertex of $C$ in a given perfect elimination ordering $v_1,\ldots,v_n$ of $G$. We claim that $C$ is equal to the closed neighborhood to the right of $v_i$, i.e., $C = N[v_i]\cap\{v_i,\ldots,v_n\}$. It is clear that $C$ is contained in $N[v_i]\cap\{v_i,\ldots,v_n\}$ since $v_i$ has no other neighbors to its right, so we indeed have $C = N[v_i]\cap\{v_i,\ldots,v_n\}$ by maximality of $C$. Therefore, there are at most $n$ maximal cliques.
\end{proof}

\begin{definition}
Let $G_1$, $G_2$ be two graphs, and suppose $C\coloneqq V(G_1)\cap V(G_2)$ is a clique in both $G_1$ and $G_2$. When we say we \textbf{glue} $G_1$ and $G_2$ together at $C$ to obtain $G$, this means $G$ is the union of $G_1$ and $G_2$: the vertex set is $V(G) = V(G_1)\cup V(G_2)$, and the edge set is $E(G) = E(G_1)\cup E(G_2)$.
\end{definition}

As is shown in~\cite{hebertjohnson2023counting}, if $G_1$ and $G_2$ are both chordal, then the resulting graph $G$ is chordal.

\subsection{Evaporation sequences}

Our algorithm for counting the number of labeled chordal graphs with a given automorphism will use the notion of evaporation sequences from~\cite{hebertjohnson2023counting}.

Suppose we are given a chordal graph $G$ and a clique $X\subseteq V(G)$. The \emph{evaporation sequence} of $G$ with \emph{exception set} $X$ is defined as follows: If $X = V(G)$, then the evaporation sequence of $G$ is the empty sequence. If $X\subsetneq V(G)$, then let $\widetilde L_1$ be the set of all simplicial vertices in $G$, and let $L_1 = \widetilde L_1\setminus X$. Suppose $L_2,\ldots,L_t$ is the evaporation sequence of $G\setminus L_1$ (with exception set $X$). Then $L_1,L_2,\ldots,L_t$ is the evaporation sequence of $G$. As is shown in~\cite{hebertjohnson2023counting}, the fact that $X$ is a clique implies that all vertices outside of $X$ eventually evaporate, so this is well-defined.

If the evaporation sequence $L_1,L_2,\ldots,L_t$ of $G$ has length $t$, then we say $G$ \emph{evaporates} at time $t$ with \emph{exception set} $X$, and $t$ is called the \emph{evaporation time}. We define $L_G(X)\coloneqq L_t$ to be the last set in the evaporation sequence of $G$, and we let $L_G(X) = \emptyset$ if the sequence is empty. Similarly, we define the evaporation time of a vertex subset. Suppose $G$ has evaporation sequence $L_1,L_2,\ldots,L_t$ with exception set $X$, and suppose $S\subseteq V(G)\setminus X$ is a nonempty vertex subset. Let $t_S$ be the largest index $i$ such that $L_i\cap S\ne\emptyset$. We say $S$ \emph{evaporates} at time $t_S$ in $G$ with exception set $X$.



\section{Sampling unlabeled chordal graphs}

In~\cite{wormald1987generating}, Wormald presented an algorithm that generates an $n$-vertex unlabeled graph uniformly at random in expected time $O(n^2)$. In this paper, we achieve a similar result for chordal graphs:

\unlabeled*

The sampling algorithm of Wormald makes use of the fact that unlabeled graphs correspond to orbits of a particular group action. Since our algorithm will follow a similar outline, we begin by discussing some of the ideas behind the algorithm of Wormald.

Suppose we have been given $n\in\N$ as input, and let $\Omega$ be the set of all labeled graphs with vertex set $[n]$. The symmetric group $S_n$ acts on $\Omega$ in the following way: For each $\pi\in S_n$ and $G\in\Omega$, $\pi\cdot G$ is the graph that results from permuting the vertex labels of $G$ according to $\pi$. The orbits of $\Omega$ under the action of $S_n$ are the isomorphism classes of labeled graphs, each of which corresponds to a unlabeled graph. Let $$\Gamma = \{(\pi,G)\in S_n\times\Omega:\pi\text{ is an automorphism of }G\}.$$ Suppose we fix an $n$-vertex unlabeled graph $H$, and let the corresponding isomorphism class of labeled graphs be $\mathcal{H}$. As is shown in~\cite{wormald1987generating}, the number of pairs $(\pi,G)\in\Gamma$ such that $G\in\mathcal{H}$ is equal to $|S_n| = n!$. This follows from the Frobenius-Burnside lemma. Therefore, if we sample a random pair $(\pi,G)\in\Gamma$ uniformly at random, and then we forget the labels on the graph $G$, this amounts to sampling an $n$-vertex unlabeled graph uniformly at random.

In the case of chordal graphs, the same statements hold true. Let $\OmegaC$ be the set of all labeled chordal graphs with vertex set $[n]$. The symmetric group $S_n$ acts on $\OmegaC$ in the same way as above, by permuting the vertex labels. The set of orbits of this group action corresponds to the set of unlabeled chordal graphs. Let $$\GammaC = \{(\pi,G)\in S_n\times\OmegaC:\pi\text{ is an automorphism of }G\}.$$ Let $H_c$ be an unlabeled chordal graph, and let $\mathcal{H}_c$ be the corresponding isomorphism class of labeled graphs. Applying the Frobenius-Burnside lemma to the orbit corresponding to $\mathcal{H}_c$ shows that the number of pairs $(\pi,G)\in\GammaC$ such that $G\in\mathcal{H}_c$ is equal to $n!$.

In~\cite{wormald1987generating}, Wormald describes an algorithm for sampling a random pair $(\pi,G)\in\Gamma$ in order to sample a random unlabeled graph. The same outline can be used to sample a random pair $(\pi,G)\in\GammaC$. However, there are two key points where some difficulty arises. First of all, in one of the steps of the algorithm that samples from $\Gamma$, it is necessary to count the number of $n$-vertex labeled graphs with a given automorphism (and sample from the set of such graphs). This is easy to do for general graphs but becomes more complicated for chordal graphs (see \cref{sec:counting_with_pi,sec:counting_with_pi_proof}). Second, the algorithm of Wormald uses the fact that the number of labeled graphs with a given automorphism $\pi$ is at most $2^{\binom{n}{2}-\mu n/2+\mu(\mu+2)/4}$, where $\mu = |M_{\pi}|$ is the number of moved points of $\pi$. To transform this into an algorithm for sampling unlabeled chordal graphs, it is necessary to prove similar bounds on the number of labeled chordal graphs with a given automorphism. These will be the bounds $B_{\mu}$ in our algorithm.

\subsection{Algorithm for sampling unlabeled chordal graphs}

Let $\Call{Count\_Chordal\_Lab}{n}$ stand for the counting algorithm in~\cite{hebertjohnson2023counting} that computes the number of $n$-vertex labeled chordal graphs. In \cref{sec:counting_with_pi,sec:counting_with_pi_proof}, we will prove the following two theorems.

\begin{restatable}{theorem}{countingLab}
\label{thm:counting_pi}
There is a deterministic algorithm that given $n\in\N$ and $\pi\in S_n$, computes the number of labeled chordal graphs with vertex set $[n]$ for which $\pi$ is an automorphism. This algorithm uses $O(2^{7\mu}n^9)$ arithmetic operations, where $\mu = |M_{\pi}|$.
\end{restatable}

\begin{restatable}{theorem}{samplingLab}
\label{thm:sampling_pi}
There is a randomized algorithm that given $n\in\N$ and $\pi\in S_n$, generates a graph uniformly at random from the set of all labeled chordal graphs with vertex set $[n]$ for which $\pi$ is an automorphism. This algorithm uses $O(2^{7\mu}n^9)$ arithmetic operations, where $\mu = |M_{\pi}|$.
\end{restatable}

In our algorithm for sampling unlabeled chordal graphs, $\Call{Count\_Chordal\_Lab}{n,\pi}$ stands for the algorithm of \cref{thm:counting_pi} and $\Call{Sample\_Chordal\_Lab}{n,\pi}$ stands for the algorithm of \cref{thm:sampling_pi}.

Recall that $[n]_0 = \{0,2,3,\ldots,n\}$. For $\mu\in[n]_0$, let $R_{\mu}\coloneqq\{\pi\in S_n:|M_{\pi}| = \mu\}$ be the set of permutations with exactly $\mu$ moved points. For $\pi\in S_n$, let $\text{Fix}(\pi)$ denote the set of $n$-vertex labeled chordal graphs $G$ such that $\pi$ is an automorphism of $G$. To follow the same approach as the algorithm of Wormald, for each $\mu\in[n]_0$, we need an upper bound $B_{\mu}$ that satisfies $B_{\mu}\ge|R_{\mu}||\text{Fix}(\pi)|$ for all $\pi\in R_{\mu}$. Let $B_0$ be the number of $n$-vertex labeled chordal graphs, which is exactly equal to $|R_0||\text{Fix}(\text{id})|$. For $2\le\mu\le\frac{n}{200\log n}$, let
\begin{align}
\label{eq:small_mu}
B_{\mu} = \left(B_0(9/10)^n+2^n2^{2n^2/9}+n\cdot 2^{n^2/4+n/2}\right)n^{\mu}\mu!,
\end{align}
and for $\frac{n}{200\log n}<\mu\le n$, let
\begin{align}
\label{eq:large_mu}
B_{\mu} = n^{2n+1}2^{n^2/4-f(\mu)}n^{\mu}\mu!,
\end{align}
where $f(\mu) = \frac{\mu^2}{900}-\frac{\mu}{10}$. In \cref{sec:correctess_unlabeled}, we will prove that we indeed have $B_{\mu}\ge|R_{\mu}||\text{Fix}(\pi)|$ for all $\pi\in R_{\mu}$, when $n$ is sufficiently large. Let $B = \sum_{\mu\in[n]_0}B_{\mu}$.



Our algorithm for sampling a random $n$-vertex unlabeled chordal graph is given in \cref{alg:unlabeled}. The general idea is as follows. For $\mu\in[n]_0$, let $\Gamma_{\mu} = \{(\pi,G)\in\GammaC:|M_{\pi}| = \mu\}$. We choose $\mu$ such that the probability of each value of $\mu$ is $B_{\mu}/B$, which is approximately equal to $\Gamma_{\mu}/\GammaC$. (We do not know how to efficiently compute the exact value of $\Gamma_{\mu}/\GammaC$ since we do not know the exact number of unlabeled chordal graphs.) Since $B_{\mu}/B$ is not exactly equal to $\Gamma_{\mu}/\GammaC$, we adjust for this by restarting with a certain probability in Step 11. We then proceed to select a random pair $(\pi,G)\in\Gamma_{\mu}$, and we output the graph $G$ without labels.

\begin{algorithm}[H]
  \caption{Unlabeled chordal graph sampler}
  \label{alg:unlabeled}
  \begin{algorithmic}[1]
    \Procedure{Sample\_Chordal\_Unlabeled}{$n$}
    \State $\triangleright$ Setup \label{op0}
    \State $B_0\gets\Call{Count\_Chordal\_Lab}{n}$
    \State Let $B_{\mu}$ be given by \cref{eq:small_mu} for $2\le\mu\le\frac{n}{200\log n}$
    \State Let $B_{\mu}$ be given by \cref{eq:large_mu} for $\frac{n}{200\log n}<\mu\le n$
    \State $B\gets\sum_{\mu\in[n]_0}B_{\mu}$
    \State
    \State $\triangleright$ Main algorithm
    \State Choose $\mu\in[n]_0$ at random such that $\mu = i$ with probability $B_i/B$ for each $i\in[n]_0$
    \State Choose $\pi\in R_{\mu}$ uniformly at random
    \State Restart (go back to Step 9) with probability
    \newline\hspace*{3em} $1-|R_{\mu}|\cdot\Call{Count\_Chordal\_Lab}{n,\pi}/B_{\mu}$
    \State $G\gets\Call{Sample\_Chordal\_Lab}{n,\pi}$
    \State Forget the labels on the vertices of $G$
    \State \Return $G$
    \EndProcedure
  \end{algorithmic}
\end{algorithm}

Step 10 can easily be implemented in $O(n)$ time in the following way. If $\mu = 0$, let $\pi = \text{id}$. For $\mu\ge 2$, we can repeatedly choose a random permutation of $[\mu]$ until we obtain a derangement. The expected number of trials for this is a constant (see~\cite{wormald1987generating}).

In Step 11, to compute $|R_{\mu}|$, we observe that $|R_0| = 1$ and $|R_{\mu}| =\,\, !\mu\binom{n}{\mu}$ for $\mu\ge 2$. Here $!\mu$ is the number of derangements of $\mu$. We compute $!\mu$ using the formula $!m = m!\sum_{i=0}^m(-1)^i/i!$ for $m\in\N$, which can be derived using the inclusion-exclusion principle.

\subsection{Correctness of \cref{alg:unlabeled}}
\label{sec:correctess_unlabeled}

See \cref{sec:counting_with_pi_proof} for the proof of correctness of the procedures $\Call{Count\_Chordal\_Lab}{n,\pi}$ and $\Call{Sample\_Chordal\_Lab}{n,\pi}$.


We need to show that the output graph is chosen uniformly at random. One ``iteration'' refers to one run of Steps 9 to 11 or 9 to 14. In a given iteration, we say the pair $(\pi,G)$ was ``chosen'' if $\pi$ was chosen from $R_{\mu}$ and $G$ was chosen by $\Call{Sample\_Chordal\_Lab}{n,\pi}$. We claim that for all $(\pi,G)\in\GammaC$, in any given iteration, the probability that $(\pi,G)$ is chosen is $1/B$. Indeed, this probability is equal to $$\frac{B_{\mu}}{B}\frac{1}{|R_{\mu}|}\frac{|R_{\mu}||\text{Fix}(\pi)|}{B_{\mu}}\frac{1}{|\text{Fix}(\pi)|} = \frac{1}{B},$$ where $\mu = |M_{\pi}|$, since the probability of choosing $G$ in Step 12 is $1/|\text{Fix}(\pi)|$. Therefore, we output all $n$-vertex unlabeled chordal graphs with equal probability.

Next, we need to show that $B_{\mu}\ge|R_{\mu}||\text{Fix}(\pi)|$ for all $\pi\in R_{\mu}$ to verify that the probability $|R_{\mu}||\text{Fix}(\pi)|/B_{\mu}$ in Step 11 is at most 1. When $\mu = 0$ this is an equality, so suppose $\mu\ge 2$. Clearly $|R_{\mu}|\le n^{\mu}\mu!$, so we just need to prove that the number of $n$-vertex labeled chordal graphs with automorphism $\pi$ is at most $B_{\mu}/(n^{\mu}\mu!)$ for all $\pi\in S_n$ with $\mu$ moved points.

In the case when $B_{\mu}$ is defined according to \cref{eq:large_mu}, this follows from \cref{theorem:overall_bound}. The proof this theorem can be found in \cref{sec:large_mu_bound}. (The bound in \cref{theorem:overall_bound} is in fact true for all values of $\mu$ --- the reason why we define $B_{\mu}$ differently for smaller values of $\mu$ will become clear when we discuss the running time in \cref{sec:running_time_unlabeled}.)

\begin{restatable}{theorem}{overallBound}
\label{theorem:overall_bound}
Let $n\in\N$, $\pi\in S_n$, and let $\mu = |M_{\pi}|$. The number of labeled chordal graphs with vertex set $[n]$ for which $\pi$ is an automorphism is at most $$n^{2n+1}2^{n^2/4-f(\mu)},$$ where $f(\mu) = \frac{\mu^2}{900}-\frac{\mu}{10}$.
\end{restatable}

For the other case, suppose $2\le\mu\le\frac{n}{200\log n}$, and suppose $\pi\in S_n$ is a permutation with $\mu$ moved points. We need to show that the number of $n$-vertex labeled chordal graphs with automorphism $\pi$ is at most $B_{\mu}/(n^{\mu}\mu!)$. We begin by reducing to the case of split graphs. A \emph{split} graph is a graph whose vertex set can be partitioned into a clique and an independent set, with arbitrary edges between the two parts. It is easy to see that every split graph is chordal. Furthermore, the following result by Bender et al.~\cite{bender1985almost} shows that a random $n$-vertex labeled chordal graph is a split graph with probability $1-o(1)$.

\begin{proposition}[Bender et al.~\cite{bender1985almost}]
\label{prop:almost_every}
If $\alpha>\sqrt 3/2$, $n$ is sufficiently large, and $G$ is a random $n$-vertex labeled chordal graph, then $$\Pr(\text{$G$ is a split graph})>1-\alpha^n.$$
\end{proposition}

Applying this proposition with $\alpha = \frac{9}{10}$ tells us that the number of $n$-vertex labeled chordal graphs that are \emph{not split} is at most $B_0\left(\frac{9}{10}\right)^n$. To bound the number of $n$-vertex labeled split graphs with automorphism $\pi$, we consider two cases: balanced split graphs and unbalanced split graphs. We say a partition of the vertex set of a split graph $G$ is a \emph{split partition} if it partitions $G$ into a clique and an independent set; i.e., a split partition is a partition that demonstrates that $G$ is split. We denote a split partition that consists of the clique $C$ and the independent set $I$ by the ordered pair $(C,I)$. We say an $n$-vertex split graph is \emph{balanced} if $|C|\ge\frac{n}{3}$ and $|I|\ge\frac{n}{3}$ for every split partition $(C,I)$ of $G$. It is easy to bound the number of unbalanced split graph as follows.

\begin{lemma}
\label{lemma:not_bal}
The number of labeled split graphs on $n$ vertices that are not balanced is at most $2^n2^{2n^2/9}$.
\end{lemma}

\begin{proof}
Suppose $(C,I)$ is a partition of $[n]$ into two parts such that $|C|<\frac{n}{3}$ or $|I|<\frac{n}{3}$. The number of labeled split graphs with this particular split partition is at most $2^{|I||C|}\le 2^{\frac{n}{3}\cdot\frac{2n}{3}} = 2^{2n^2/9}$. Therefore, the number of unbalanced labeled split graphs on $n$ vertices is at most $2^n2^{2n^2/9}$ since there are at most $2^n$ possible partitions.
\end{proof}

The following lemma will be useful for bounding the number of balanced split graphs.

\begin{lemma}
\label{lemma:invariant_split}
Let $\pi\in S_n$, and let $G$ be an $n$-vertex labeled split graph. If $\pi$ is an automorphism of $G$, then there exists a split partition $(C,I)$ of $G$ such that $C$ and $I$ are invariant under $\pi$.
\end{lemma}

\begin{proof}
Let $\hat C$ be the set of vertices that belong to the clique in every split partition of $G$, let $\hat I$ be the set of vertices that belong to the independent set in every split partition of $G$, and let $\hat Q = V(G)\setminus(\hat C\cup\hat I)$. By Observation 7.3 in~\cite{hebertjohnson2023counting}, every vertex in $\hat Q$ is adjacent to every vertex in $\hat C$ and is non-adjacent to every vertex in $\hat I$. By Lemma 7.4 in~\cite{hebertjohnson2023counting}, $\hat Q$ is either a clique or an independent set. Suppose $\pi$ is an automorphism of $G$. If $\hat Q$ is a clique, then the split partition $(\hat C\cup\hat Q,\hat I)$ has the desired property. If $\hat Q$ is an independent set, then the split partition $(\hat C,\hat I\cup\hat Q)$ has the desired property.
\end{proof}

\begin{lemma}
\label{lemma:bal}
Let $\pi\in S_n$, and suppose $|M_{\pi}|\ge 2$. The number of balanced labeled split graphs $G$ on $n$ vertices such that $\pi$ is an automorphism of $G$ is at most $$n\cdot 2^{n^2/4+n/2}.$$
\end{lemma}

\begin{proof}
Suppose $(C,I)$ is a partition of $[n]$ into two parts. Let $i = |I|$ and $c = |C|$. The number of labeled split graphs with this particular split partition is $2^{ic}$. Therefore, the number of labeled split graphs with automorphism $\pi$ for which $(C,I)$ has the property from \cref{lemma:invariant_split} is at most $2^{ic}$. Let $Z_{(C,I)}$ be the number of such graphs. Whenever two vertices $u,v\in I$ (resp.\ $C$) belong to the same cycle in the cycle decomposition of $\pi$, $u$ and $v$ must have the same relationship as each other to each of the vertices in $C$ (resp.\ $I$). Therefore, we in fact have an upper bound of $\max\{2^{(i-1)c},2^{i(c-1)}\}$ on $Z_{(C,I)}$, since $\pi$ has at least two moved points. If $i\ge\frac{n}{3}$ and $c\ge\frac{n}{3}$, then we have $\max\{2^{(i-1)c},2^{i(c-1)}\}\le 2^{n^2/4-n/2}$.

By \cref{lemma:invariant_split}, every balanced labeled split graph on $n$ vertices with automorphism $\pi$ has a split partition $(C,I)$ such that $C$ and $I$ are invariant under $\pi$. Furthermore, this split partition is balanced (i.e., both parts have size at least $\frac{n}{3}$). Therefore, the number of balanced labeled split graphs on $n$ vertices with automorphism $\pi$ is at most $$\sum_{\lceil\frac{n}{3}\rceil\le i\le\lfloor\frac{2n}{3}\rfloor}2^n\cdot2^{n^2/4-n/2}\le n\cdot 2^{n^2/4+n/2}$$ since the number of partitions $(C,I)$ with $|I| = i$ is certainly at most $2^n$.
\end{proof}

Putting together \cref{prop:almost_every,lemma:not_bal,lemma:bal}, we can see that $$\left(B_0(9/10)^n+2^n2^{2n^2/9}+n\cdot 2^{n^2/4+n/2}\right)n^{\mu}\mu!$$ is an upper bound on $|R_{\mu}||\text{Fix}(\pi)|$ when $n$ is sufficiently large.

Let $N_0$ be the cutoff such that this works for $n\ge N_0$. To be precise, when implementing this algorithm, we would solve the problem by brute force if $n<N_0$ (by generating all possible $n$-vertex chordal graphs and then selecting one at random), and we would run \cref{alg:unlabeled} as written if $n\ge N_0$.

\subsection{Running time of \cref{alg:unlabeled}}
\label{sec:running_time_unlabeled}

The running time of Steps 1-6 is $O(n^7)$ arithmetic operations since that is the running time of $\Call{Count\_Chordal\_Lab}{n}$. In this section, we will show that the rest of the algorithm uses only $O(n^7)$ arithmetic operations in expectation.

If we choose $\mu = 0$ in Step 9 of \cref{alg:unlabeled}, then the algorithm is guaranteed to terminate in that iteration. Thus the expected number of iterations is at most $B/B_0$. Let $T$ be the expected running time of \cref{alg:unlabeled} after completing Step 6 (this is where we choose $\mu$ at random and the loop begins). In \cref{lemma:overall_running_time}, we show that $ \E[T]$ is at most the product of $B/B_0$ and the expected running time of one iteration.

Let $N$ be the number of iterations of \cref{alg:unlabeled}, and let $T_j$ be the time spent in iteration~$j$ for each $j\in[N]$. We have $T = \sum_{j=1}^N T_j$.

\begin{lemma}
\label{lemma:overall_running_time}
We have $\E[T]\le\frac{B}{B_0}\E[T_1]$.
\end{lemma}

\begin{proof}
Clearly $\E[T]$ is finite since the expected number of iterations is finite and the procedures $\Call{Count\_Chordal\_Lab}{n,\pi}$ and $\Call{Sample\_Chordal\_Lab}{n,\pi}$ have worst-case running time bounds. Therefore, we can solve for $\E[T]$ in the following way.

The steps that we run in one iteration (Steps 9-14) do not depend on $j$ --- they are always the same, regardless of how many iterations have happened so far. Thus we have $\E\left[\sum_{j=2}^N T_j\,\big|\,N>1\right] = \E[T]$, which implies $\E[T\mid N>1] = \E[T_1\mid N>1]+\E[T]$. Therefore, we have
\begin{align*}
\E[T] &= \E[T\mid N=1]\Pr(N=1)+\E[T\mid N>1]\Pr(N>1) \\
&= \E[T_1\mid N=1]\Pr(N=1)+\big(\E[T_1\mid N>1]+\E[T]\big)\cdot\Pr(N>1)
\end{align*}
\begin{align*}
\implies\E[T](1-\Pr(N>1)) &= \E[T_1\mid N=1]\Pr(N=1)+\E[T_1\mid N>1]\Pr(N>1) \\
&= \E[T_1]
\end{align*}
\begin{align*}
\implies\E[T] &= \frac{\E[T_1]}{\Pr(N=1)}\le\frac{B}{B_0}\E[T_1].
\end{align*}
\end{proof}


For $\mu\in[n]_0$, let $T(n,\mu)$ be an upper bound on the time it takes to run one iteration, assuming we have chosen this particular value of $\mu$ in Step 9. By the running time of $\Call{Count\_Chordal\_Lab}{n,\pi}$ and $\Call{Sample\_Chordal\_Lab}{n,\pi}$, we can assume $T(n,\mu) = O(2^{7\mu}n^9)$. We have $$\E[T_1] = \sum_{\mu\in[n]_0}\frac{B_{\mu}}{B}T(n,\mu).$$ To prove a bound on $\E[T_1]$, we will start by proving a bound on $B_{\mu}/B$ for all $\mu\in[n]_0$. Since $B_0\le B$, it is sufficient to prove a bound on $B_{\mu}/B_0$ for all $\mu\in[n]_0$. The following lemma gives us a lower bound on $B_0$.

\begin{lemma}
\label{lemma:B_0}
For $n\ge 2$, the number of $n$-vertex labeled chordal graphs is at least $$\frac{2^n2^{n^2/4}}{n^2}.$$
\end{lemma}

\begin{proof}
It is enough to just consider $n$-vertex labeled split graphs with a split partition $(C,I)$ such that $|C| = \lfloor\frac{n}{2}\rfloor$. Since $\binom{n}{\lfloor\frac{n}{2}\rfloor}\ge 2^n/n$ for $n\ge 2$, the number of such graphs is at least $$\binom{n}{\lfloor\frac{n}{2}\rfloor}\frac{2^{n^2/4}}{n}\ge\frac{2^n2^{n^2/4}}{n^2}.$$ We divide by $n$ in the first expression since each split graph can have up to $n$ distinct split partitions in which $C$ is of a given size~\cite{bina2015note}.
\end{proof}

\begin{lemma}
\label{lemma:ratio_small_mu}
Suppose $n\ge 13$. If $2\le\mu\le\frac{n}{200\log n}$, then $$\frac{B_{\mu}}{B_0}\le\frac{3}{n^{16\mu}}.$$
\end{lemma}

\begin{proof}
Let $B_{\mu}^{(1)}$, $B_{\mu}^{(2)}$, and $B_{\mu}^{(3)}$ be the three terms that are added together in \cref{eq:small_mu}, in order, so that $B_{\mu} = \left(B_{\mu}^{(1)}+B_{\mu}^{(2)}+B_{\mu}^{(3)}\right)n^{\mu}\mu!$. We have $$\frac{B_{\mu}^{(1)}n^{\mu}\mu!}{B_0} = \left(\frac{9}{10}\right)^n n^{\mu}\mu!,$$ and we claim that this is at most $1/n^{16\mu}$. Since $\mu!\le n^{\mu}$, it is sufficient to show $n^{18\mu}\le(10/9)^n$, which is true when $\mu\le\frac{n}{200\log n}$.

For the next term, by \cref{lemma:B_0} we have $$\frac{B_{\mu}^{(2)}n^{\mu}\mu!}{B_0}\le n^2 2^{-n^2/36}n^{\mu}\mu!.$$ To see that this is at most $1/n^{16\mu}$, it is sufficient to show $n^{19/200}\le 2^{n/36}$ since $\mu\le\frac{n}{200}$. This is indeed true for $n\ge 13$.

For the third term, by \cref{lemma:B_0} we have $$\frac{B_{\mu}^{(3)}n^{\mu}\mu!}{B_0}\le n^3 2^{-n/2}n^{\mu}\mu!.$$ To see that this is at most $1/n^{16\mu}$, it is sufficient to show $n^{20\mu}\le 2^{n/2}$, which is true when $\mu\le\frac{n}{40\log n}$. Adding together these three terms, we obtain $B_{\mu}/B_0\le 3/n^{16\mu}$.
\end{proof}

\begin{lemma}
\label{lemma:ratio_large_mu}
For sufficiently large $n$, if $\frac{n}{200\log n}<\mu\le n$, then $$\frac{B_{\mu}}{B_0}\le\frac{1}{n^{16\mu}}.$$
\end{lemma}

\begin{proof}
\cref{lemma:B_0} implies $B_0\ge 2^{n^2/4}$ for $n\ge 4$, so we have $$\frac{B_{\mu}}{B_0}\le n^{2n+1}2^{-f(\mu)}n^{\mu}\mu!,$$ where $f(\mu) = \frac{\mu^2}{900}-\frac{\mu}{10}$. We claim that this expression is at most $1/n^{16\mu}$. Since $\mu!\le n^{\mu}$, it is sufficient to show $n^{2n+1}n^{18\mu}\le 2^{f(\mu)}$. When $n$ is sufficiently large,\footnote{This holds for $n\ge 2.6\cdot 10^5$.} we have $\log^3 n\le\frac{1}{200^2\cdot 1800}\frac{n^2}{2n+1}$. Since $\mu\ge\frac{n}{200\log n}$, this implies
\begin{align}
\label{ineq:1}
n^{2n+1}\le 2^{\mu^2/1800}.
\end{align}
When $n$ is sufficiently large,\footnote{This holds for $n\ge 3.3\cdot 10^9$.} we also have $n\ge 18\cdot 900\cdot 200\log^2 n+90\cdot 200\log n$. Since $\mu\ge\frac{n}{200\log n}$, this implies
\begin{align}
\label{ineq:2}
n^{18\mu}\le 2^{\mu^2/1800-\mu/10}.
\end{align}
Multiplying \cref{ineq:1,ineq:2} gives us the desired bound of $n^{2n+1}n^{18\mu}\le 2^{f(\mu)}$.

\end{proof}

By \cref{lemma:ratio_small_mu,lemma:ratio_large_mu}, the above summation for $\E[T_1]$ is at most $T(n,0)+O(1)$ since $T(n,\mu)$ is certainly at most $O(n^{16\mu})$. For an iteration in which we have chosen $\mu = 0$, when counting and sampling $n$-vertex labeled chordal graphs with automorphism $\pi = \text{id}$, we can simply run the counting and sampling algorithms of~\cite{hebertjohnson2023counting}, rather than passing in $\pi = \text{id}$ as an input. Thus the expected running time $\E[T_1]$ of one iteration is at most $O(n^7)$ arithmetic operations.
\cref{lemma:ratio_small_mu,lemma:ratio_large_mu} also immediately give us a bound on $B/B_0$, since we have $$\frac{B}{B_0} = \frac{B_0}{B_0}+\frac{B_2}{B_0}+\ldots+\frac{B_n}{B_0} = O(1).$$ Therefore, by \cref{lemma:overall_running_time}, the overall running time is at most $O(n^7)$ arithmetic operations in expectation.


\section{Counting labeled chordal graphs with a given automorphism}
\label{sec:counting_with_pi}


In this section, we describe the algorithm for $\Call{Count\_Chordal\_Lab}{n,\pi}$, which counts the number of labeled chordal graphs with a given automorphism. This is the algorithm of \cref{thm:counting_pi}. In \cref{sec:counting_with_pi_proof} we will prove correctness, analyze the running time, and derive the corresponding sampling algorithm (\cref{thm:sampling_pi}).




\countingLab*

There is a known dynamic-programming algorithm for computing the number $n$-vertex labeled chordal graphs that uses $O(n^7)$ arithmetic operations, if we do not require the graphs to have a particular automorphism~\cite{hebertjohnson2023counting}. Our algorithm is closely based on that one, but we add more arguments and more details to each of the recurrences to keep track of the behavior of the automorphism. As was done in~\cite{hebertjohnson2023counting}, we evaluate the recurrences top-down using memoization.

\subsection{Reducing from counting chordal graphs to counting connected chordal graphs}

For $k\in\N$, let $a(k)$ denote the number of labeled chordal graphs with vertex set $[k]$, and let $c(k)$ denote the number of \emph{connected} labeled chordal graphs with vertex set $[k]$. The algorithm of~\cite{hebertjohnson2023counting} begins by reducing from counting chordal graphs to counting connected chordal graphs via the following recurrence, which appears as Lemma 3.13 in~\cite{hebertjohnson2023counting}. (We omit the ``$\omega$-colorable'' requirement since we will not need that here.)

\begin{lemma}[\cite{hebertjohnson2023counting}]
\label{lemma:disconn}
The number of labeled chordal graphs with vertex set $[k]$ is given by $$a(k) = \sum_{k'=1}^k\binom{k-1}{k'-1}c(k')a(k-k')$$ for all $k\in\N$.
\end{lemma}

Here $k'$ stands for the number of vertices in the connected component that contains the label 1. The remaining connected components have a total of $k-k'$ vertices. Since this recurrence is relatively simple, most of the difficulty in the algorithm of~\cite{hebertjohnson2023counting} lies in the recurrences for counting \emph{connected} chordal graphs. However, when counting graphs with a given automorphism, the step of reducing to connected graphs is already quite a bit more involved.

Suppose we are given $n\in\N$ and $\pi\in S_n$ as input. From now on, whenever we refer to $n$ or $\pi$, we mean these particular values from the input.

We will first define $a(k,p,M)$ and $c(k,p,M)$, which are variations of $a(k)$ and $c(k)$ that only count graphs for which $\pi^p|_{V(G)}$ is an automorphism (see \cref{def:a_and_c}). Here $\pi^p$ stands for $\pi$ to the power $p$, i.e., the permutation that arises from applying $\pi$ a total of $p$ times. The reason for raising $\pi$ to a power will be apparent in the recurrence for $a(k,p,M)$. In the initial, highest-level recursive call, we will have $p = 1$ and thus $\pi^p = \pi$.

In~\cite{hebertjohnson2023counting}, when counting the number of possibilities for a subgraph of size $k'$ (for example, a connected component), the authors essentially relabel that subgraph to have vertex set $[k']$, so that one can count the number of possibilities using, for example, $c(k')$. In our algorithm, we want to relabel the vertices of each subgraph in a similar way. However, this time, we do not relabel the vertices that are moved by $\pi$. This ensures that the automorphism in each later recursive call will still be $\pi$, or a permutation closely related to $\pi$.

As a consequence, the vertex sets of the subgraphs that we wish to count become slightly more complicated. For example, suppose we wish to count the number of possibilities for a 5-vertex subgraph that originally contains the moved vertices $2,8,9\in M_{\pi}$. Since we do not relabel the moved vertices, the resulting vertex set after relabeling is $\{1,2,3,8,9\}$ rather than $\{1,2,3,4,5\}$. More formally, suppose we have been given $k\in\N$ and $M\subseteq M_{\pi}$, where $|M|\le k$. Let $V$ be the set of the first $k-|M|$ natural numbers in $\N\setminus M_{\pi}$. We define $V_{k,M}\coloneqq V\cup M$. For example, if $k = 5$ and $M = M_{\pi} = \{2,8,9\}$, then $V_{k,M} = \{1,2,3,8,9\}$. Intuitively, $V_{k,m}$ is the label set of size $k$ whose intersection with $M_{\pi}$ is $M$ and that otherwise contains labels that are as small as possible.

For $\hat\pi\in S_n$ and $M\subseteq[n]$, recall that $M$ is \emph{invariant} under $\hat\pi$ if $\hat\pi(i)\in M$ for all $i\in M$. For $M',M\subseteq[n]$, we write $M'\subseteq_{\hat\pi}M$ to indicate that $M'\subseteq M$ and $M'$ is invariant under $\hat\pi$. Intuitively, $M'$ is a subset of $M$ that respects the cycles of $\hat\pi$ by taking all or nothing of each cycle.

\begin{definition}
\label{def:a_and_c}
Suppose $k,p\in[n]$ and suppose $M\subseteq_{\pi^p}M_{\pi}$, where $|M|\le k$. Let $a(k,p,M)$ (resp.\ $c(k,p,M)$) denote the number of labeled chordal graphs (resp.\ \textbf{connected} labeled chordal graphs) with vertex set $V_{k,M}$ for which $\pi^p|_{V(G)}$ is an automorphism.
\end{definition}

Our algorithm returns $a(n,1,M_{\pi})$. This is precisely the number of labeled chordal graphs with vertex set $[n]$ and automorphism $\pi$ since $V_{n,M_{\pi}} = [n]$.

Suppose we have been given fixed values of the arguments $k,p,M$ of $a(k,p,M)$. Let $s$ be the smallest label in the vertex set $V_{k,M}$. For $k'\in[k]$, let $\mathcal{P}_{k'}$ be the family of sets $M'\subseteq_{\pi^p}M$ such that $|M|-k+k'\le|M'|\le k'$ and such that the following condition holds: if $s\in M$, then $s\in M'$, and otherwise, $|M'|\le k'-1$. This last condition will ensure that $s$ belongs to the connected component of size $k'$ in the recurrence for $a(k,p,M)$.

Suppose $C\subseteq[n]$, $\sigma\in S_n$. For an element $i\in C$, we say the \emph{period} of $i$ with respect to $(C,\sigma)$ is the smallest positive integer $j$ such that $\sigma^j(i)\in C$. Let $\mathcal{Q}$ be the family of sets $C\subseteq M$ such that all elements of $C$ have the same period $j\ge 2$ with respect to $(C,\pi^p)$, and such that $s\in C$. For a set $C\in\mathcal{Q}$, we write $p_C$ to denote the period of the elements of $C$ (with respect to $(C,\pi^p)$), and we let $C_{\sigma}\coloneqq C\cup\sigma(C)\cup\cdots\cup\sigma^{p_C-1}(C)$ denote the union of the sets that $C$ is mapped to by powers of $\sigma$, where $\sigma = \pi^p$.

To compute $a(k,p,M)$, we use the following recurrence. The dot in front of the curly braces denotes multiplication. Note that we have $|M'|\le k'$ and $|M\setminus M'|\le k-k'$ by the definition of $\mathcal{P}_{k'}$.

\begin{restatable}{lemma}{aPi}
\label{lemma:a_pi}
Let $k,p\in[n]$ and suppose $M\subseteq_{\pi^p}M_{\pi}$, where $|M|\le k$. We have
\begin{align*}
a(k,p,M) &= \sum_{\substack{1\le k'\le k \\ M'\in\mathcal{P}_{k'}}}\hspace{-0.2em}c(k',p,M')a(k-k',p,M\setminus M')\cdot
\begin{cases}
\binom{k-|M|}{k'-|M'|} & \text{ if $s\in M$} \\
\binom{k-1-|M|}{k'-1-|M'|} & \text{ otherwise}
\end{cases} \\
&\quad+\sum_{C\in\mathcal{Q}}c(|C|,p\cdot p_C,C)a(k-p_C |C|,p,M\setminus C_{\pi^p}).
\end{align*}
\end{restatable}

For some intuition, suppose $G$ is a graph counted by $a(k,p,M)$, and let $C$ be the connected component of $G$ that contains $s$. The first line of the recurrence for $a(k,p,M)$ covers the case when $C$ is invariant under $\pi^p$. This case is analogous to \cref{lemma:disconn}. In the first summation, $k'$ stands for $|C|$ and $M'$ stands for the set of vertices in $C$ that can be moved by $\pi^p$. If $s\in M$, then in addition to the vertices in $M'$, we need to choose $k'-|M'|$ additional vertices for $C$ so that $|C| = k'$. For these, we must choose non-moved vertices, so there are $k-|M|$ possible vertices to choose from. If $s\notin M$, then we subtract 1 from each of the numbers in the binomial coefficient since we already know that $s$ is a non-moved vertex in $C$.

The second line covers the case when $C$ is not invariant under $\pi^p$, which means all of $C$ is mapped to some other connected component of $G$ by $\pi^p$. In this case, we have $p_C$ components of size $|C|$ that are all isomorphic to $C$, and the rest of the graph has $k-p_C|C|$ vertices. We do not need a binomial coefficient in this case because all of the vertices in $C$ are moved. There are $c(|C|,p\cdot p_C,C)$ possibilities for the edges of $C$ since $\pi^{p\cdot p_C}$ is an automorphism of $C$. As an example, suppose $p = 1$. If we apply $\pi^p = \pi$ to the vertices of $C$ a total of $p_C$ times, then the image $\pi^{p_C}(C)$ is equal to $C$, although these two sets might not match up pointwise. To ensure that the image $\pi^{p_C}(C)$ matches up with the edges of $C$, we require that $\pi^{p_C}$ is an automorphism of $C$. This is why we need the argument $p$ in $a(k,p,M)$ and $c(k,p,M)$.

Note that for a graph $G$ counted by $a(k,p,M)$, we have $M_{\pi^p}\cap V(G)\subseteq M_{\pi}\cap V(G)\subseteq M$ since $V(G) = V_{k,M}$. This means no vertex in $V(G)\setminus M$ is moved by $\pi^p$, so we are free to relabel these vertices in the proof of \cref{lemma:a_pi} without changing the automorphism. In the initial recursive call $a(n,1,M_{\pi})$, we have $p = 1$ and $M = M_{\pi}$, so $M_{\pi^p}\cap V(G) = M$. Later on in the algorithm, it is possible that some vertices in $M$ are not actually moved by $\pi^p$. For example, in the previous paragraph, it could happen that $\pi^{p\cdot p_C}$ is the identity on $C$ (and then $p\cdot p_C$ becomes the new value of $p$). However, we always have $M\subseteq M_{\pi}$ by the definition of $a(k,p,M)$, so if $|M_{\pi}| = \mu$, then $|M|\le \mu$. This fact will be useful for the running time analysis in \cref{sec:running_time_pi}.

The proof of \cref{lemma:a_pi}, along with the proofs of all of the following recurrences, can be found in \cref{sec:counting_with_pi_proof}.

\subsection{Recurrences for counting connected chordal graphs}

To compute $c(k,p,M)$, we will define various counter functions that are analogous to those in~\cite{hebertjohnson2023counting}. First, we recall the counter functions from~\cite{hebertjohnson2023counting}. We refer to functions 1-4 in \cref{def:original_counters} as $g$-functions, and we refer to the others as $f$-functions. See Figure 1 in~\cite{hebertjohnson2023counting} for an illustration of all of these functions.

\begin{definition}[\cite{hebertjohnson2023counting}]
\label{def:original_counters}
The following functions count various subclasses of chordal graphs. The arguments $t,x,l,k,z$ are nonnegative integers, where $t\le n$, $z\le n$, $x+k\le n$ for $g$-functions, and $x+l+k\le n$ for $f$-functions. These also satisfy the domain requirements listed below.
\begin{enumerate}
    \item $g(t,x,k,z)$ is the number of labeled connected chordal graphs $G$ with vertex set $[x+k]$ that evaporate in time at most $t$ with exception set $X\coloneqq[x]$, where $X$ is a clique, with the following property: every connected component of $G\setminus X$ (if any) has at least one neighbor in $X\setminus[z]$. \textbf{Domain:} $t\ge 0$, $x\ge 1$, $z<x$.
    \item $\tilde g(t,x,k,z)$ is the same as $g(t,x,k,z)$, except every connected component of $G\setminus X$ (if any) evaporates at time exactly $t$ in $G$. \emph{Note: A graph with $V(G) = X$ would be counted because in that case, $\tilde g$ is the same as $g$.}
    \textbf{Domain:} $t\ge 1$, $x\ge 1$, $z<x$.
    \item $\tilde g_p(t,x,k,z)$ is the same as $\tilde g(t,x,k,z)$, except no connected component of $G\setminus X$ sees all of~$X$. \textbf{Domain:} $t\ge 1$, $x\ge 1$, $z<x$.
    \item $\tilde g_1(t,x,k)$ and $\tilde g_{\ge 2}(t,x,k)$ are the same as $\tilde g(t,x,k,z)$, except every connected component of $G\setminus X$ sees all of $X$ (hence we no longer require every component of $G\setminus X$ to have a neighbor in $X\setminus[z]$), and furthermore, for $\tilde g_1$ we require that $G\setminus X$ has exactly one connected component, and for $\tilde g_{\ge 2}$ we require that $G\setminus X$ has at least two components. \textbf{Domain for $\tilde g_1$:} $t\ge 1$, $x\ge 0$. \textbf{Domain for $\tilde g_{\ge 2}$:} $t\ge 1$, $x\ge 1$.
    \item $f(t,x,l,k)$ is the number of labeled connected chordal graphs $G$ with vertex set $[x+l+k]$ that evaporate at time exactly $t$ with exception set $X\coloneqq[x]$, such that $G\setminus X$ is connected, $L_G(X) = \{x+1,\ldots,x+l\}$, and $X\cup L_G(X)$ is a clique. \textbf{Domain:} $t\ge 1$, $x\ge 0$, $l\ge 1$.
    \item $\tilde f(t,x,l,k)$ is the same as $f(t,x,l,k)$, except every connected component of $G\setminus(X\cup L_G(X))$ evaporates at time exactly $t-1$ in $G$, and there exists at least one such component, i.e., $X\cup L_G(X)\subsetneq V(G)$.
    \textbf{Domain:} $t\ge 2$, $x\ge 0$, $l\ge 1$.
    \item $\tilde f_p(t,x,l,k)$ is the same as $\tilde f(t,x,l,k)$, except no connected component of $G\setminus(X\cup L_G(X))$ sees all of $X\cup L_G(X)$. \textbf{Domain:} $t\ge 2$, $x\ge 0$, $l\ge 1$.
    \item $\tilde f_p(t,x,l,k,z)$ is the same as $\tilde f_p(t,x,l,k)$, except rather than requiring that $G\setminus X$ is connected, we require that $G\setminus[z]$ is connected. \textbf{Domain:} $t\ge 2$, $x\ge 0$, $l\ge 1$, $z\le x$.
\end{enumerate}
\end{definition}

The counter functions for our algorithm will be similar to these, except we only want to count graphs with a particular automorphism. Therefore, we will have several more arguments in addition to the usual ones $t,x,l,k,z$ from \cref{def:original_counters}. As above, the argument $p$ will indicate that $\pi^p$ is the current automorphism. We also introduce the arguments $M_X$, $M_L$, $M_Z$, and $M_K$, each of which is a subset of $M_{\pi}$. Roughly, these sets specify which vertices --- in $X$, $L\coloneqq L_G(X)$, $Z\coloneqq[z]$, and the rest of the graph, respectively --- can be moved by the current permutation (but the sets $X$, $L$, and $Z$ will be modified slightly).

In \cref{def:original_counters}, the $g$-functions count graphs with vertex set $[x+k]$, and the $f$-functions count graphs with vertex set $[x+l+k]$. For our algorithm, we will make some adjustments to these vertex sets to ensure that the vertices moved by the current permutation still appear in the graph. This is similar to how we defined $V_{k,M}$ above. To do so, we define several symbols for these vertex sets and vertex subsets ($V_{args}$, etc.). Each of these depends on the list of arguments of the function in question, which we denote by $args$. For example, when defining $g$, we have $args = \begin{pmatrix}
t\, & x & k & z \\
p\, & M_X & M_K & M_Z
\end{pmatrix}$ (see \cref{def:new_counters}).

First, suppose $args$ comes from one of the $g$-functions in \cref{def:new_counters}. Let $V_X$ be the set of the first $x-|M_X|$ natural numbers in $\N\setminus M_{\pi}$, and let $V_K$ be the set of the first $k-|M_K|$ natural numbers in $\N\setminus(V_X\cup M_{\pi})$. We define $X_{args}\coloneqq V_X\cup M_X$ and $V_{args}\coloneqq X_{args}\cup V_K\cup M_K$. Also, if $z$ is included in $args$, then let $V_Z$ be the set of the first $z-|M_Z|$ natural numbers in $\N\setminus M_{\pi}$. In this case, we define $Z_{args}\coloneqq V_Z\cup M_Z$.

For the other case, suppose $args$ comes from one of the $f$-functions. Let $V_X$ be the set of the first $x-|M_X|$ natural numbers in $\N\setminus M_{\pi}$, let $V_L$ be the set of the first $l-|M_L|$ natural numbers in $\N\setminus(V_X\cup M_{\pi})$, and let $V_K$ be the set of the first $k-|M_K|$ natural numbers in $\N\setminus(V_X\cup V_L\cup M_{\pi})$. We define $X_{args}\coloneqq V_X\cup M_X$, $L_{args} = V_L\cup M_L$, and $V_{args}\coloneqq X_{args}\cup L_{args}\cup V_K\cup M_K$. Also, if $z$ is included in $args$, then let $V_Z$ be the set of the first $z-|M_Z|$ natural numbers in $\N\setminus M_{\pi}$. In this case, we again define $Z_{args}\coloneqq V_Z\cup M_Z$.

These vertex subsets still have the same sizes as they did in the original algorithm: the size of $V_{args}$ is $x+k$ or $x+l+k$, $|X_{args}| = x$, $|L_{args}| = l$, the rest of the graph has size $k$, and $|Z_{args}| = z$. Just as we had $Z\subseteq X$ in the original algorithm, we now have $Z_{args}\subseteq X_{args}$. This is because we require $M_Z\subseteq M_X$ in \cref{def:new_counters}, and we also require $z-|M_Z|\le x-|M_X|$, which implies $V_Z\subseteq V_X$.

For $g$-functions, we have $M_{\pi^p}\cap V_{args}\subseteq M_X\cup M_K$, and for $f$-functions, we have $M_{\pi^p}\cap V_{args}\subseteq M_X\cup M_L\cup M_K$, by the definition of $V_{args}$. We are now ready to define our new counter functions.

\begin{definition}
\label{def:new_counters}
The following functions count various subclasses of chordal graphs. The arguments $t,x,l,k,z$ are nonnegative integers with the same domains as in \cref{def:original_counters}. We have $p\in[n]$, and the arguments $M_X,M_L,M_K,M_Z$ are subsets of $M_{\pi}$. All other requirements for their domains are specified below.
\begin{enumerate}
    \item $\g{t}{x}{k}{z}{p}{M_X}{M_K}{M_Z}$, $\gTilde{t}{x}{k}{z}{p}{M_X}{M_K}{M_Z}$, $\gTildeP{t}{x}{k}{z}{p}{M_X}{M_K}{M_Z}$, \\
    $\gTildeOne{t}{x}{k}{p}{M_X}{M_K}$, and $\gTildeTwo{t}{x}{k}{p}{M_X}{M_K}$ are the same as $g(t,x,k,z)$, $\tilde g(t,x,k,z)$, $\tilde g_p(t,x,k,z)$, $\tilde g_1(t,x,k)$, and $\tilde g_{\ge 2}(t,x,k)$, respectively, except we only count graphs for which $\pi^p|_{V(G)}$ is an automorphism, and we make the following changes to the vertices of the graph: the vertex set is $V_{args}$ rather than $[x+k]$, the exception set is $X_{args}$ rather than $[x]$, and $[z]$ is replaced by $Z_{args}$.
        
    \smallskip
    \textbf{Domain:} $M_X\subseteq_{\pi^p}M_{\pi}$, $M_K\subseteq_{\pi^p}M_{\pi}\setminus M_X$, $M_Z\subseteq_{\pi^p}M_X$, $|M_X|\le x$, $|M_K|\le k$, $|M_Z|\le z$, and $z-|M_Z|\le x-|M_X|$.
    
    \medskip
    \item $\f{t}{x}{l}{k}{p}{M_X}{M_L}{M_K}$, $\fTilde{t}{x}{l}{k}{p}{M_X}{M_L}{M_K}$, $\fTildeP{t}{x}{l}{k}{p}{M_X}{M_L}{M_K}$, and \\ $\fTildePWithZ{t}{x}{l}{k}{z}{p}{M_X}{M_L}{M_K}{M_Z}$ are the same as $f(t,x,l,k)$, $\tilde f(t,x,l,k)$, $\tilde f_p(t,x,l,k)$, and $\tilde f_p(t,x,l,k,z)$, respectively, except we only count graphs for which $\pi^p|_{V(G)}$ is an automorphism, and we make the following changes to the vertices of the graph: the vertex set is $V_{args}$ rather than $[x+l+k]$, the exception set is $X_{args}$ rather than $[x]$, the last set to evaporate is $L_{args}$ rather than $\{x+1,\ldots,x+l\}$, and $[z]$ is replaced by $Z_{args}$.
    
    \smallskip
    \textbf{Domain:} $M_X\subseteq_{\pi^p}M_{\pi}$, $M_L\subseteq_{\pi^p}M_{\pi}\setminus M_X$, $M_K\subseteq_{\pi^p}M_{\pi}\setminus(M_X\cup M_L)$, $M_Z\subseteq_{\pi^p}M_X$, $|M_X|\le x$, $|M_L|\le l$, $|M_K|\le k$, $|M_Z|\le z$, and $z-|M_Z|\le x-|M_X|$.
\end{enumerate}
\end{definition}

For a graph $G$ counted by one of these functions, $\pi^p|_{V(G)}$ is indeed a permutation of $V(G)$ since $M_X$ and $M_K$ (and $M_L$ if needed) are invariant under $\pi^p$.

To compute $c(k,p,M)$, we consider all possibilities for the evaporation time. In the following recurrence, $\tilde g_1$ (with those particular arguments) counts the number of labeled connected chordal graphs with vertex set $V_{args} = V_{k,M}$ and automorphism $\pi^p|_{V(G)}$ that evaporate at time exactly $t$ with empty exception set.

\begin{restatable}{lemma}{cPi}
Let $k,p\in[n]$ and suppose $M\subseteq_{\pi^p}M_{\pi}$, where $|M|\le k$. We have $$c(k,p,M) = \sum_{t=1}^k\gTildeOne{t}{0}{k}{p}{\emptyset}{M}.$$
\end{restatable}

To compute $\tilde g_1$, we consider all possibilities for the size $l$ of $L_{args}$. The set $M$ in the inner sum stands for the set of vertices in $L_{args}$ that can be moved by $\pi^p$. Note that we have $|M|\le l$ and $|M_K\setminus M|\le k-l$ by the definition of $\mathcal{I}_l$. For $\tilde g_1$ and all of the following functions, we recommend reading the analogous recurrences in~\cite{hebertjohnson2023counting} for further insight into what is happening with the arguments $t,x,l,k,z$ in each recursive call.

\begin{restatable}{lemma}{gTildeOnePi}
For $\tilde g_{1}$, we have $$\gTildeOne{t}{x}{k}{p}{M_X}{M_K} = \sum_{l=1}^k\sum_{M\in\mathcal{I}_l}\binom{k-|M_K|}{l-|M|}\f{t}{x}{l}{k-l}{p}{M_X}{M}{M_K\setminus M},$$ where $\mathcal{I}_l = \{M\subseteq_{\pi^p}M_K:|M_K|-k+l\le|M|\le l\}$.
\end{restatable}

To compute $f$, we consider all possibilities for the set of labels that appear in connected components of $G\setminus(X_{args}\cup L_{args})$ that evaporate at time exactly $t-1$. These components correspond to the recursive call to $\tilde f$. The set $M$ stands for the set of vertices in components that evaporate at time exactly $t-1$ that can be moved by $\pi^p$. Note that we have $|M|\le k'$ and $|M_K\setminus M|\le k-k'$ by the definition of $\mathcal{I}_{k'}$. Since $|M_L|\le l$, we also have $x-|M_X|\le x+l-|M_X\cup M_L|$, which is required by the domain of $g$.


\begin{restatable}{lemma}{fPi}
\label{lemma:f_pi}
For $f$, we have
\begin{align*}
&\f{t}{x}{l}{k}{p}{M_X}{M_L}{M_K} = \\
&\sum_{k'=1}^k\sum_{M\in\mathcal{I}_{k'}}\binom{k-|M_K|}{k'-|M|}\fTilde{t}{x}{l}{k'}{p}{M_X}{M_L}{M}\g{t-2}{x+l}{k-k'}{x}{p}{M_X\cup M_L}{M_K\setminus M}{M_X},
\end{align*}
where $\mathcal{I}_{k'} = \{M\subseteq_{\pi^p}M_K:|M_K|-k+k'\le|M|\le k'\}$.
\end{restatable}

To compute $g$, we consider all possibilities for the set of labels that appear in connected components of $G\setminus X_{args}$ that evaporate at time exactly $t$. These components correspond to the recursive call to $\tilde g$. The set $M$ stands for the set of vertices in components that evaporate at time exactly $t$ that can be moved by $\pi^p$.

\begin{restatable}{lemma}{gPi}
\label{lemma:g_pi}
For $g$, we have
\begin{align*}
&\g{t}{x}{k}{z}{p}{M_X}{M_K}{M_Z} = \\
&\sum_{k'=0}^k\sum_{M\in\mathcal{I}_{k'}}\binom{k-|M_K|}{k'-|M|}\gTilde{t}{x}{k'}{z}{p}{M_X}{M}{M_Z}\g{t-1}{x}{k-k'}{z}{p}{M_X}{M_K\setminus M}{M_Z},
\end{align*}
where $\mathcal{I}_{k'} = \{M\subseteq_{\pi^p}M_K:|M_K|-k+k'\le|M|\le k'\}$.
\end{restatable}

For the next recurrence, we will need a few more definitions. Suppose we have been given fixed values of the arguments of $\tilde g$. Let $s$ be the smallest label in $V_{args}\setminus X_{args}$. For $k'\in[k]$, let $\mathcal{P}_{k'}$ be the family of sets $M\subseteq_{\pi^p}M_K$ such that $|M_K|-k+k'\le|M|\le k'$ and such that the following condition holds: if $s\in M_K$, then $s\in M$, and otherwise, $|M|\le k'-1$. For $x'\in[x]$, let $\mathcal{I}_{x'} = \{M'\subseteq_{\pi^p}M_X:|M'|\le x'\}$.

Let $\mathcal{Q}$ be the family of pairs $(C,M')$, where $C\subseteq M_K$ and $M'\subseteq M_X$, such that all elements of $C$ have the same period $p_C\ge 2$ with respect to $(C,\pi^p)$, such that $s\in C$, and such that $M'$ is invariant under $\pi^{p\cdot p_C}$. For a set $C$ from such a pair, we let $C_{\sigma}\coloneqq C\cup\sigma(C)\cup\cdots\cup\sigma^{p_C-1}(C)$, where $\sigma = \pi^p$.

To compute $\tilde g$, we consider all possibilities for the label set of the connected component $C$ of $G\setminus X_{args}$ that contains $s$. In the definition of $\tilde g$, the components of $G\setminus X_{args}$ are similar enough (since they all evaporate at time $t$) that they can potentially be mapped to one another by $\pi^p$. Thus we consider two cases: either $C$ is invariant under $\pi^p$, or $C$ is mapped to some other component of $G\setminus X_{args}$ by $\pi^p$. We add together the summations from these two cases, in a similar way to the recurrence for $a(k,p,M)$. In the recurrence for $\tilde g$, $k'$ stands for $|C|$, and $x'$ stands for $|N(C)|$. The set $M$ stands for the set of vertices in $C$ that can be moved by $\pi^p$, and $M'$ stands for the set of vertices in $N(C)$ that can be moved by $\pi^p$.

\begin{restatable}{lemma}{gTildePi}
\label{lemma:g_tilde_pi}
For $\tilde g$, we have
\begin{align*}
&\gTilde{t}{x}{k}{z}{p}{M_X}{M_K}{M_Z} = \sum_{\substack{1\le k'\le k \\ 1\le x'\le x \\ M\in\mathcal{P}_{k'} \\ M'\in\mathcal{I}_{x'}}}\gTildeOne{t}{x'}{k'}{p}{M'}{M}\gTilde{t}{x}{k-k'}{z}{p}{M_X}{M_K\setminus M}{M_Z}
\end{align*}
\vspace{-1.3cm}
\begin{align*}
\qquad\qquad\qquad\qquad\qquad\qquad\qquad\qquad&\cdot
\begin{cases}
\binom{k-|M_K|}{k'-|M|} & \text{ if $s\in M_K$} \\
\binom{k-1-|M_K|}{k'-1-|M|} & \text{ otherwise}
\end{cases} \\
&\cdot
\begin{cases}
\binom{x-|M_X|}{x'-|M'|} & \text{ if $M'\not\subseteq M_Z$} \\
\binom{x-|M_X|}{x'-|M'|}-\binom{z-|M_Z|}{x'-|M'|} & \text{ otherwise}
\end{cases}
\end{align*}
\vspace{-0.2cm}
\begin{align*}
\qquad\qquad\qquad\qquad\qquad+\sum_{\substack{1\le x'\le x \\ (C,M')\in\mathcal{Q}}}\gTildeOne{t}{x'}{|C|}{p\cdot p_C}{M'}{C}\gTilde{t}{x}{k-p_C|C|}{z}{p}{M_X}{M_K\setminus C_{\pi^p}}{M_Z}
\end{align*}
\vspace{-0.8cm}
\begin{align*}
\qquad\qquad\qquad\qquad\qquad\qquad\qquad\qquad&\cdot
\begin{cases}
\binom{x-|M_X|}{x'-|M'|} & \text{ if $M'\not\subseteq M_Z$} \\
\binom{x-|M_X|}{x'-|M'|}-\binom{z-|M_Z|}{x'-|M'|} & \text{ otherwise}.
\end{cases}
\end{align*}
\end{restatable}



To compute $\tilde f$, we consider three cases: either no component of $G\setminus(X_{args}\cup L_{args})$ sees all of $X_{args}\cup L_{args}$, exactly one component sees all of $X_{args}\cup L_{args}$, or at least two components see all of $X_{args}\cup L_{args}$. The recursive calls to $\tilde g_1$ and $\tilde g_{\ge 2}$ correspond to the component/components that see all of $X_{args}\cup L_{args}$. In the second and third cases, the set $M$ stands for the set of vertices that can be moved by $\pi^p$ in components that see all of $X_{args}\cup L_{args}$.

\begin{restatable}{lemma}{fTildePi}
For $\tilde f$, we have
\begin{align*}
\fTilde{t}{x}{l}{k}{p}{M_X}{M_L}{M_K} &= \fTildeP{t}{x}{l}{k}{p}{M_X}{M_L}{M_K} \\
&\hspace{-10em}+\sum_{\substack{1\le k'\le k \\ M\in\mathcal{I}_{k'}}}\binom{k-|M_K|}{k'-|M|}\gTildeOne{t-1}{x+l}{k'}{p}{M_X\cup M_L}{M}\fTildeP{t}{x}{l}{k-k'}{p}{M_X}{M_L}{M_K\setminus M} \\
&\hspace{-12em}+\sum_{\substack{1\le k'\le k \\ M\in\mathcal{I}_{k'}}}\binom{k-|M_K|}{k'-|M|}\gTildeTwo{t-1}{x+l}{k'}{p}{M_X\cup M_L}{M}\gTildeP{t-1}{x+l}{k-k'}{x}{p}{M_X\cup M_L}{M_K\setminus M}{M_X},
\end{align*}
where $\mathcal{I}_{k'} = \{M\subseteq_{\pi^p}M_K:|M_K|-k+k'\le|M|\le k'\}$.
\end{restatable}

For the next recurrence, suppose we have been given fixed values of the arguments of $\tilde g_{\ge 2}$. Let $s$ be the smallest label in $V_{args}\setminus X_{args}$, and let $\mathcal{P}_{k'}$ be defined as it was in the recurrence for $\tilde g$. Let $\mathcal{Q}$ be the family of sets $C\subseteq M$ such that all elements of $C$ have the same period $p_C\ge 2$ with respect to $(C,\pi^p)$, and such that $s\in C$. For a set $C\in\mathcal{Q}$, we let $C_{\sigma}\coloneqq C\cup\sigma(C)\cup\cdots\cup\sigma^{p_C-1}(C)$, where $\sigma = \pi^p$.

To compute $\tilde g_{\ge 2}$, we consider all possibilities for the label set of the connected component $C$ of $G\setminus X_{args}$ that contains $s$. This is somewhat similar to the recurrence for $\tilde g$, but this time we do not need to keep track of the neighborhood of $C$, since every component of $G\setminus X_{args}$ sees all of $X_{args}$.

\begin{restatable}{lemma}{gTildeTwoPi}
For $\tilde g_{\ge 2}$, we have
\begin{align*}
&\gTildeTwo{t}{x}{k}{p}{M_X}{M_K} = \\
&\sum_{\substack{1\le k'\le k \\ M\in\mathcal{P}_{k'}}}\gTildeOne{t}{x}{k'}{p}{M_X}{M}\Bigg(\gTildeOne{t}{x}{k-k'}{p}{M_X}{M_K\setminus M}+\,\gTildeTwo{t}{x}{k-k'}{p}{M_X}{M_K\setminus M}\Bigg)
\end{align*}
\vspace{-0.7cm}
\begin{align*}
\qquad\qquad\qquad\qquad\qquad\qquad\qquad\qquad&\cdot
\begin{cases}
\binom{k-|M_K|}{k'-|M|} & \text{ if $s\in M_K$} \\
\binom{k-1-|M_K|}{k'-1-|M|} & \text{ otherwise}
\end{cases}
\end{align*}
\vspace{-0.3cm}
\begin{align*}
\qquad\hspace{-2.5em}+\sum_{C\in\mathcal{Q}}\gTildeOne{t}{x}{|C|}{p\cdot p_C}{M_X}{C}\Bigg(\gTildeOne{t}{x}{k-p_C|C|}{p}{M_X}{M_K\setminus C^{\pi}}+\,\gTildeTwo{t}{x}{k-p_C|C|}{p}{M_X}{M_K\setminus C^{\pi}}\Bigg).
\end{align*}
\end{restatable}

To compute $\tilde g_p$, all we need to do is make a slight adjustment to the recurrence for $\tilde g$.

\begin{restatable}{lemma}{gTildePPi}
The recurrence for $\tilde g_p$ is exactly the same as the recurrence for $\tilde g$ in \cref{lemma:g_tilde_pi}, except for the two sums over $x'$: In both summations, rather than summing over $x'$ such that $1\le x'\le x$, we sum over $x'$ such that $1\le x'\le x-1$.
\end{restatable}

To compute $\tilde f_p$, we first observe that when $z = x$, requiring $G\setminus Z_{args}$ to be connected is the same as requiring $G\setminus X_{args}$ to be connected.

\begin{restatable}{lemma}{fTildePPi}
We have $\fTildeP{t}{x}{l}{k}{p}{M_X}{M_L}{M_K} = \fTildePWithZ{t}{x}{l}{k}{x}{p}{M_X}{M_L}{M_K}{M_X}$.
\end{restatable}

For the next $\tilde f_p$ recurrence, we need one last round of definitions. Suppose we have been given fixed values of the arguments of $\tilde f_p$, including $z$ and $M_Z$. Let $s$ be the smallest label in $V_{args}\setminus(X_{args}\cup L_{args})$. For $k'\in[k]$, let $\mathcal{P}_{k'}$ be the family of sets $M\subseteq_{\pi^p}M_K$ such that $|M_K|-k+k'\le|M|\le k'$ and such that the following condition holds: if $s\in M_K$, then $s\in M$, and otherwise, $|M|\le k'-1$. For $0\le x'\le x$, let $\mathcal{I}_{x'} = \{M'\subseteq_{\pi^p}M_X:|M'|\le x'\}$. Also, for $0\le l'\le l$, let $\mathcal{J}_{l'} = \{M'\subseteq_{\pi^p}M_L:|M'|\le l'\}$.

Let $\mathcal{Q}$ be the family of triples $(C,M_X',M_L')$, where $C\subseteq M_K$, $M_X'\subseteq M_X$, and $M_L'\subseteq M_L$, such that all elements of $C$ have the same period $p_C\ge 2$ with respect to $(C,\pi^p)$, such that $s\in C$, and such that $M_X'\cup M_L'$ is invariant under $\pi^{p\cdot p_C}$. For a set $C$ from such a triple, we let $C_{\sigma}\coloneqq C\cup\sigma(C)\cup\cdots\cup\sigma^{p_C-1}(C)$, where $\sigma = \pi^p$.

To compute $\tilde f_p$ with the argument $z$, we consider all possibilities for the label set of the component $C$ of $G\setminus(X_{args}\cup L_{args})$ that contains $s$. There are two possible cases: either $C$ is invariant under $\pi^p$, or $C$ is mapped to some other component of $G\setminus(X_{args}\cup L_{args})$ by $\pi^p$. We add together the summations from these two cases, both of which are now quite large. In the following recurrence, $k'$ stands for $|C|$, $x'$ stands for $|N(C)\cap X_{args}|$, and $l'$ stands for $|N(C)\cap L_{args}|$. For the moved vertices, $M$ stands for the set of vertices in $C$ that can be moved by $\pi^p$, $M_X'$ stands for the set of vertices in $N(C)\cap X_{args}$ that can be moved by $\pi^p$, and $M_L'$ stands for the set of vertices in $N(C)\cap L_{args}$ that can be moved by $\pi^p$.




\begin{restatable}{lemma}{fTildePWithZPi}
For $\tilde f_p$ with the argument $z$, we have
\begin{align*}
&\fTildePWithZ{t}{x}{l}{k}{z}{p}{M_X}{M_L}{M_K}{M_Z} = \sum_{\substack{1\le k'\le k \\ 0\le x'\le x \\ 0\le l'\le l \\ 0<x'+l'<x+l}}\sum_{\substack{M\in\mathcal{P}_{k'} \\ M_
X'\in\mathcal{I}_{x'} \\ M_
L'\in\mathcal{J}_{l'}}}\gTildeOne{t-1}{x'+l'}{k'}{p}{M_X'\cup M_L'}{M}
\end{align*}
\vspace{-0.4cm}
\begin{align*}
\quad\qquad\qquad\qquad\qquad&\cdot\binom{l-|M_L|}{l'-|M_L'|}\cdot
\begin{cases}
\binom{k-|M_K|}{k'-|M|} & \text{ if $s\in M_K$} \\
\binom{k-1-|M_K|}{k'-1-|M|} & \text{ otherwise}
\end{cases} \\
&\cdot
\begin{cases}
\binom{x-|M_X|}{x'-|M_X'|} & \text{ if $l'>0$ or $M_X'\not\subseteq M_Z$} \\
\binom{x-|M_X|}{x'-|M_X'|}-\binom{z-|M_Z|}{x'-|M_X'|} & \text{ otherwise}
\end{cases} \\
&\cdot
\begin{cases}
\fTildePWithZ{t}{x+l'}{l-l'}{k-k'}{z}{p}{M_X\cup M_L'}{M_L\setminus M_L'}{M_K\setminus M}{M_Z} & \text{ if $l'<l$} \\[12pt]
\gTildeP{t-1}{x+l}{k-k'}{z}{p}{M_X\cup M_L'}{M_K\setminus M}{M_Z} & \text{ otherwise}
\end{cases}
\end{align*}
\vspace{-0.3cm}
\begin{align*}
\quad+\sum_{\substack{0\le x'\le x \\ 0\le l'\le l \\ 0<x'+l'<x+l \\ (C,M_X',M_L')\in\mathcal{Q}}}\hspace{-1.8em}\gTildeOne{t-1}{x'+l'}{|C|}{p\cdot p_C}{M_X'\cup M_L'}{C}\cdot\binom{l-|M_L|}{l'-|M_L'|}
\end{align*}
\vspace{-1.35cm}
\begin{align*}
\quad\qquad\qquad\qquad\qquad&\cdot
\begin{cases}
\binom{x-|M_X|}{x'-|M_X'|} & \text{ if $l'>0$ or $M_X'\not\subseteq M_Z$} \\
\binom{x-|M_X|}{x'-|M_X'|}-\binom{z-|M_Z|}{x'-|M_X'|} & \text{ otherwise}
\end{cases} \\
&\cdot
\begin{cases}
\fTildePWithZ{t}{x+l'}{l-l'}{k-p_C|C|}{z}{p}{M_X\cup M_L'}{M_L\setminus M_L'}{M_K\setminus C^{\pi}}{M_Z} & \text{ if $l'<l$} \\[12pt]
\gTildeP{t-1}{x+l}{k-p_C|C|}{z}{p}{M_X\cup M_L'}{M_K\setminus C^{\pi}}{M_Z} & \text{ otherwise}.
\end{cases}
\end{align*}
\end{restatable}

The base cases for all of these counter functions are the same as in~\cite{hebertjohnson2023counting} since they only depend on the arguments $t,x,l,k,z$.



\section{Proofs of \cref{thm:counting_pi,thm:sampling_pi}}
\label{sec:counting_with_pi_proof}

In this section, we prove correctness of the recurrences from \cref{sec:counting_with_pi} and complete the proofs of \cref{thm:counting_pi,thm:sampling_pi}.

\subsection{Automorphism properties}
\label{sec:auto_properties}

First of all, we show that automorphisms preserve evaporation time.

\begin{lemma}
\label{lemma:invariant_evap_time}
Suppose $G$ is a labeled chordal graph that evaporates at time $t$ with exception set $X$, where $X\subseteq V(G)$ is a clique, and let $L_1,\ldots,L_t$ be the evaporation sequence. If $\hat\pi$ is an automorphism of $G$ such that $X$ is invariant under $\hat\pi$, then $L_i$ is invariant under $\hat\pi$ for all $i\in[t]$.
\end{lemma}

\begin{proof}
We proceed by induction on $i$. Let $\widetilde L_1$ be the set of all simplicial vertices in $G$. If the neighborhood of a vertex $v\in V(G)$ is a clique, then the neighborhood of $\hat\pi(v)$ must also be a clique, so $\widetilde L_1$ is invariant under $\hat\pi$. Since $X$ is also invariant under $\hat\pi$, this implies that $L_1 = \widetilde L_1\setminus X$ is invariant under $\hat\pi$.

For the induction step, suppose $L_1,\ldots,L_i$ are all invariant under $\hat\pi$ for some $i\in[t-1]$. This means $\hat\pi|_{V(G)\setminus(L_1\cup\cdots\cup L_i)}$ is an automorphism of $G\setminus(L_1\cup\cdots\cup L_i)$. By the same argument as in the base case, the set $\widetilde L_{i+1}$ of simplicial vertices of $G\setminus(L_1\cup\cdots\cup L_i)$ is invariant under $\hat\pi|_{V(G)\setminus(L_1\cup\cdots\cup L_i)}$ and also under $\hat\pi$. Hence $L_{i+1} = \widetilde L_{i+1}\setminus X$ is invariant under $\hat\pi$ as well.
\end{proof}

Next, we prove two lemmas that relate the automorphisms of a graph $G$ to the automorphisms of its induced subgraphs.

\begin{lemma}
\label{lemma:subgraph_auto}
Let $G$ be a labeled chordal graph with an automorphism $\hat\pi$ and induced subgraphs $G_1$ and $G_2$ such that $V(G_1)\cup V(G_2) = V(G)$. If $X\coloneqq V(G_1)\cap V(G_2)$ and $A\coloneqq V(G_1)\setminus V(G_2)$ are both invariant under $\hat\pi$, then $\hat\pi|_{V(G_1)}$ is an automorphism of $G_1$ and $\hat\pi|_{V(G_2)}$ is an automorphism of $G_2$.
\end{lemma}

\begin{proof}
We know $\hat\pi|_{V(G_1)}$ is a bijection from $V(G_1)$ to $V(G_1)$ since $X$ and $A$ are invariant under $\hat\pi$. Moreover, this map is in fact an automorphism of $G_1$ since $\hat\pi$ is an automorphism of $G$ and $G_1$ is an induced subgraph. Similarly, $\hat\pi|_{V(G_2)}$ is a bijection from $V(G_2)$ to $V(G_2)$ since $X$ and $V(G_2)\setminus V(G_1)$ are invariant under $\hat\pi$, and this map is in fact an automorphism $G_2$.
\end{proof}

\begin{lemma}
\label{lemma:supergraph_auto}
Let $G$ be a labeled chordal graph with induced subgraphs $G_1$ and $G_2$ such that $V(G_1)\cup V(G_2) = V(G)$ and such that there are no edges from $V(G_1)\setminus V(G_2)$ to $V(G_2)\setminus V(G_1)$. If $\hat\pi$ is a permutation of $V(G)$ such that $\hat\pi|_{V(G_1)}$ is an automorphism of $G_1$ and $\hat\pi|_{V(G_2)}$ is an automorphism of $G_2$, then $\hat\pi$ is an automorphism of $G$.
\end{lemma}

\begin{proof}
Let $u,v\in V(G)$. If $u$ and $v$ both belong to $V(G_1)$, then $u$ and $v$ are adjacent in $G_1$ if and only if $\hat\pi(u)$ and $\hat\pi(v)$ are adjacent in $G_1$ since $\hat\pi|_{V(G_1)}$ is an automorphism of $G_1$. Since $G_1$ is an induced subgraph of $G$, this means $u$ and $v$ are adjacent in $G$ if and only if $\hat\pi(u)$ and $\hat\pi(v)$ are adjacent in $G$. The same statement also holds if $u,v\in V(G_2)$. The only remaining case is when one vertex is in $V(G_1)\setminus V(G_2)$ and the other is in $V(G_2)\setminus V(G_1)$. In this case, $u$ and $v$ are not adjacent in $G$, and neither are $\hat\pi(u)$ and $\hat\pi(v)$ since $V(G_1)\cap V(G_2)$ is invariant under $\hat\pi$.
\end{proof}

\subsection{Proofs of recurrences}
\label{sec:recurrence_proofs_pi}

Recall that we are given $n$ and $\pi$ as input. The proofs of all of these recurrences build upon the proofs of the analogous recurrences in~\cite{hebertjohnson2023counting}. For this reason, we highly recommend reading the proof of correctness of the counting algorithm in~\cite{hebertjohnson2023counting} before reading this section in detail. In the proof of almost every recurrence here, we will cite the analogous lemma from~\cite{hebertjohnson2023counting} to prove various properties such as the evaporation time of each subgraph.

\begin{remark}
\label{remark:injective}
One important piece of the original proofs in~\cite{hebertjohnson2023counting} is that for each recurrence, the authors describe two injective functions. For example, in the proof of the recurrence for $g(t,x,k,z)$, there is an injective function that maps each graph $G$ counted by $g(t,x,k,z)$ to a tuple $(G_1,G_2,A)$, as well as an injective function that maps each tuple $(G_1,G_2,A)$ to such a graph $G$. In this paper, we only prove injectivity in detail in the proof of the recurrence for $\tilde g$, since this statement for each of the other recurrences is either similar to $\tilde g$ or simpler than $\tilde g$. This can be found in \cref{lemma:injective_le,lemma:injective_ge}, after the recurrence for $\tilde g$.
\end{remark}

We use uppercase letters to denote the set of graphs counted by each of the counter functions. For example, the set of graphs counted by $a(k,p,M)$ is denoted by $A(k,p,M)$, the set of graphs counted by $\g{t}{x}{k}{z}{p}{M_X}{M_K}{M_Z}$ is denoted by $\G{t}{x}{k}{z}{p}{M_X}{M_K}{M_Z}$, and so on. We also use similar notation for the original counter functions: $G(t,x,k,z)$ is the set of graphs counted by  $g(t,x,k,z)$, $\widetilde G(t,x,k,z)$ is the set of graphs counted by $\tilde g(t,x,k,z)$, and so on. We proceed in the order in which our counter functions were defined, so that similar functions and proofs appear consecutively. (This is somewhat different from the ordering of the lemma numbers.)

\aPi*

\begin{proof}
Let $\mathcal{G}_{inv}$ be the set of graphs $G\in A(k,p,M)$ such that $C$ is invariant under $\pi^p$, where $C$ is the connected component of $G$ that contains $s$, and let $\mathcal{G}_{move} = A(k,p,M)\setminus\mathcal{G}_{inv}$. To prove this recurrence, we will first describe a map from $\mathcal{G}_{inv}$ to the set of tuples $(G_1,G_2,k',M',\hat C)$ with the properties given below, and then we will describe a map from $\mathcal{G}_{move}$ to the set of tuples $(G_1,G',C)$ given below. This will imply that $a(k,p,M)$ is at most the summation above. Next, to show that $a(k,p,M)$ is at least the summation above, we will describe a map from the set of tuples $(G_1,G_2,k',M',\hat C)$ to $\mathcal{G}_{inv}$, as well as a map from the set of tuples $(G_1,G',C)$ to $\mathcal{G}_{move}$. Strictly speaking, to prove these two bounds ($\le$ and $\ge$), it remains to show that all of these maps are injective. See \cref{remark:injective}.

To see that $a(k,p,M)$ is at most the summation above, suppose $G\in A(k,p,M)$. Let $C$ be the set of vertices in the connected component of $G$ that contains $s$. There are two possible cases: either $C$ is invariant under $\pi^p$, or all of $C$ is mapped to some other component of $G$ by $\pi^p$. If we are in the first case, then let $G_1 = G[C]$ and $G_2 = G\setminus C$. Let $k' = |C|$, $M' = C\cap M$, $\hat C = C\setminus M'$, and $D = V(G_2)$. Note that $|\hat C| = k'-|M'|$. Since $\pi^p$ maps $C$ to itself, this means $\pi^p|_{V(G_1)}$ is an automorphism of $G_1$, $\pi^p|_{V(G_2)}$ is an automorphism of $G_2$, and we have $M'\subseteq_{\pi^p}M$. Clearly $|M'|\le k'$. We also have $k'-|M'|\le k-|M|$ since the number of vertices that are fixed points of $\pi^p$ in $G_1$ is at most the number of fixed points in $G$. If $s\in M$, then we know $s\in M'$ by the definition of $C$. Otherwise, if $s\notin M$, then we have $|M'|\le k'-1$ since $s\in C\setminus M'$. Hence $M'\in\mathcal{P}_{k'}$.

Now relabel $G_1$ by applying $\phi(\hat C,V_{k',M'}\setminus M')$ to the labels in $\hat C$, and relabel $G_2$ by applying $\phi(D\setminus M,V_{k-k',M\setminus M'}\setminus M)$ to the labels in $D\setminus M$. We have $M_{\pi^p}\cap V_{k,M}\subseteq M$ by the definition of $V_{k,M}$, so by only relabeling vertices outside of $M$, we know that we have not relabeled any vertices that are moved by $\pi^p$. Thus $\pi^p|_{V(G_1)}$ and $\pi^p|_{V(G_2)}$ continue to be automorphisms of $G_1$ and $G_2$. Clearly $D\setminus M = D\setminus(M\setminus M')$. We also have $V_{k-k',M\setminus M'}\setminus M = V_{k-k',M\setminus M'}\setminus(M\setminus M')$ since $V_{k-k',M\setminus M'}\setminus(M\setminus M')$ is disjoint from $M_{\pi}$ and is therefore disjoint from $M$. Hence $G_1$ now has the vertex set $V_{k',M'}$ and $G_2$ has the vertex set $V_{k-k',M\setminus M'}$. Since $G_1$ is connected, we have $G_1\in C(k',p,M')$ and $G_2\in A(k-k',p,M\setminus M')$.

If we are in the second case, where $C$ is mapped to some other component of $G$ by $\pi^p$, then we have $C\in\mathcal{Q}$ since every connected component of $G$ is mapped to a connected component of $G$ by $\pi^p$. Let $C_1 = C$, and let $C_i = \pi^p(C_{i-1})$ for $2\le i\le p_C$. Let $G_i = G[C_i]$ for $i\in[p_C]$. The graphs $G_2,\ldots,G_{p_C}$ are all isomorphic to $G_1$ since $\pi^p,\ldots,\pi^{p(p_C-1)}$ are automorphisms of $G$. (This ensures that the map that takes $G$ to $(G_1,G',C)$ is injective, where $G'$ is as defined below.)

For the remainder of the graph, let $D = V(G)\setminus(C_1\cup\cdots\cup C_{p_C})$, and let $G' = G[D]$. Leave $G_1$ as is, and relabel $G'$ by applying $\phi(D\setminus M,V_{\hat k,\hat M}\setminus M)$ to the labels in $D\setminus M$, where $\hat k = k-p_C |C|$ and $\hat M = M\setminus C_{\pi^p}$. We know $\pi^{p\cdot p_C}|_{V(G_1)}$ is an automorphism of $G_1$ since $\pi^{p\cdot p_C}|_{V(G)}$ is an automorphism of $G$ that maps $C$ to itself. We also know $\pi^p|_{V(G')}$ is an automorphism of $G'$, so $G_1\in C(|C|,p\cdot p_C,C)$ and $G'\in A(k-p_C |C|,p,M\setminus C_{\pi^p})$. Hence $a(k,p,M)$ is at most the summation above.

We now show that $a(k,p,M)$ is bounded below by this summation. For the first case, suppose we are given $1\le k'\le k$, a set $M'\in\mathcal{P}_{k'}$, a graph $G_1\in C(k',p,M')$, a graph $G_2\in A(k-k',p,M\setminus M')$, and a subset $\hat C\subseteq V_{k,M}\setminus M$ of size $k'-|M'|$ that contains $s$ if $s\notin M$. Let $C = M'\cup\hat C$, and let $D = V_{k,M}\setminus C$. We construct a graph $G$ as follows: Relabel $G_1$ by applying $\phi(V_{k',M'}\setminus M',\hat C)$ to the labels in $V_{k',M'}\setminus M'$, and relabel $G_2$ by applying $\phi(V_{k-k',M\setminus M'}\setminus M,D\setminus M)$ to the labels in $V_{k-k',M\setminus M'}\setminus M$. Now let $G$ be the union of $G_1$ and $G_2$ (by taking the union of the vertex sets and the edge sets). Since $\pi^p|_{V(G_1)}$ is an automorphism of $G_1$ and $\pi^p|_{V(G_2)}$ is an automorphism of $G_2$, this means $\pi^p|_{V(G)}$ is an automorphism of $G$, so we have $G\in A(k,p,M)$.

For the second case, suppose we are given a set $C\in\mathcal{Q}$, a graph $G_1\in C(|C|,p\cdot p_C,C)$, and a graph $G'\in A(k-p_C |C|,p,M\setminus C_{\pi^p})$. Let $G_2,\ldots,G_{p_C}$ all be equal to $G_1$ initially. Leave $G_1$ as is, and relabel $G_i$ by applying $\pi^{p(i-1)}$ to its label set for all $2\le i\le p_C$. Let $D = V_{k,M}\setminus C_{\pi^p}$. Relabel $G'$ by applying $\phi(V_{\hat k,\hat M}\setminus M,D\setminus M)$ to the labels in $V_{\hat k,\hat M}\setminus M$, where $\hat k = k-p_C |C|$ and $\hat M = M\setminus C_{\pi^p}$. Finally, let $G$ be the union of $G_1$, …, $G_{p_C}$, and $G'$. The label sets of $G_1,\ldots,G_{p_C}$ are all pairwise disjoint since $C\in\mathcal{Q}$. If we apply $\pi^p$ to all of the labels in $G_i$ for some $1\le i<p_C$, then the resulting graph is $G_{i+1}$. Furthermore, if we apply $\pi^p$ to the labels in $G_{p_C}$, then the resulting graph is the same as the graph we obtain if we apply $\pi^{p\cdot p_C}$ to the labels in $G_1$. Since $\pi^{p\cdot p_C}|_{V(G_1)}$ is an automorphism of $G_1$, this graph is in fact equal to $G_1$. This means $\pi^p|_{V(\hat G)}$ is an automorphism of $\hat G$, where $\hat G$ is the union of $G_1,\ldots,G_{p_C}$. Since we also know $\pi^p|_{V(G')}$ is an automorphism of $G'$, this implies that $\pi^p|_{V(G)}$ is an automorphism of $G$. Therefore, $G\in A(k,p,M)$, so $a(k,p,M)$ is at least the summation above.
\end{proof}

\cPi*

\begin{proof}
This is similar to the proof of the recurrence for $c(n)$ in~\cite{hebertjohnson2023counting} (see the beginning of Section 5.3 in~\cite{hebertjohnson2023counting}). The only difference is that in this recurrence, both $c(k,p,M)$ and $\tilde g_1$ are counting graphs $G$ with vertex set $V_{args} = V_{k,M}$ for which $\pi^p|_{V(G)}$ is an automorphism.
\end{proof}

\gPi*

\begin{proof}
Let $\mathcal{G} = \G{t}{x}{k}{z}{p}{M_X}{M_K}{M_Z}$. To prove this recurrence, we will first describe a map from $\mathcal{G}$ to the set of tuples $(G_1,G_2,k',M,\hat A)$ with the properties given below, which will imply that $g$ is at most the summation above. Next, to show that $g$ is at least the summation above, we will describe a map from the set of tuples $(G_1,G_2,k',M,\hat A)$ to $\mathcal{G}$. Strictly speaking, to prove these two bounds ($\le$ and $\ge$), it remains to show that both of these maps are injective. See \cref{remark:injective}.

Let $args$, $args_1$, and $args_2$ be the arguments of $g$ on the left, $\tilde g$, and $g$ on the right, respectively. To see that $g$ is at most the summation above, suppose $G\in\G{t}{x}{k}{z}{p}{M_X}{M_K}{M_Z}.$ Let $A$ be the set of vertices in components $C$ of $G\setminus X_{args}$ such that $C$ evaporates at time exactly $t$ in $G$ with exception set $X_{args}$, and let $B$ be the set of vertices of all other components of $G\setminus X_{args}$. Let $G_1 = G[X_{args}\cup A]$, $G_2 = G[X_{args}\cup B]$, $k' = |A|$, $M$ = $A\cap M_K$, and $\hat A = A\setminus M$. Note that $|\hat A| = k' - |M|$. We know $X_{args}$ is invariant under $\pi^p$ since $M_X$ is invariant under $\pi^p$. If $\pi^p$ were to map some vertex in $A$, which belongs to a component $C$ of $G\setminus X_{args}$, to a vertex in $B$, then $\pi^p$ would also map all of $C$ to $B$, including the vertices of $C$ that evaporate at time $t$. By \cref{lemma:invariant_evap_time}, this is impossible, so $A$ is invariant under $\pi^p$, and so is $M$. Since $G$ has the automorphism $\pi^p|_{V(G)}$, by \cref{lemma:subgraph_auto} this means $G_1$ has the automorphism $\pi^p|_{V(G_1)}$ and $G_2$ has the automorphism $\pi^p|_{V(G_2)}$. Clearly $|M|\le k'$, and we also have $k'-|M|\le k-|M_K|$ since the number of vertices that are fixed points of $\pi^p$ in $A$ is at most the number of fixed points in $G\setminus X_{args}$. Hence $M\in\mathcal{I}_{k'}$.

Now relabel $G_1$ by applying $\phi(X_{args}\setminus M_X,X_{args_1}\setminus M_X)$ to the labels in $X_{args}\setminus M_X$ and applying $\phi(\hat A,V_{args_1}\setminus(X_{args_1}\cup M))$ to the labels in $\hat A$. Relabel $G_2$ by applying $\phi(X_{args}\setminus M_X,X_{args_2}\setminus M_X)$ to the labels in $X_{args}\setminus M_X$ and applying $\phi(B\setminus M_K,V_{args_2}\setminus(X_{args_2}\cup M_K))$ to the labels in $B\setminus M_K$. We have only relabeled vertices that are not moved by $\pi^p$, so $G_1$ and $G_2$ still have the automorphisms $\pi^p|_{V(G_1)}$ and $\pi^p|_{V(G_2)}$. By the proof of Lemma 3.6 in~\cite{hebertjohnson2023counting}, we have $G_1\in\widetilde G(t,x,k',z)$ and $G_2\in G(t-1,x,k-k',z)$.\footnote{\label{footnote:technically}Technically, this is not quite true since graphs in $\widetilde G(t,x,k',z)$ and $G(t-1,x,k-k',z)$ have vertex set $[x+k']$ and $[x+k-k']$, respectively. However, suppose we relabel $G_1$ by applying $\phi(Z_{args_1},[z])$ to the labels in $Z_{args_1}$, applying $\phi(X_{args_1}\setminus Z_{args_1},[x]\setminus[z])$ to the labels in $X_{args_1}\setminus Z_{args_1}$, and applying $\phi(V_{args_1}\setminus X_{args_1},[x+k']\setminus[x])$ to the labels in $V_{args_1}\setminus X_{args_1}$. After relabeling $G_1$ in this way, we have $G_1\in\widetilde G(t,x,k',z)$. Also, if we relabel $G_2$ in a similar way, then we have $G_2\in G(t-1,x,k-k',z)$.} Therefore, we have $G_1\in\GTilde{t}{x}{k'}{z}{p}{M_X}{M}{M_Z}$ and $G_2\in\G{t-1}{x}{k-k'}{z}{p}{M_X}{M_K\setminus M}{M_Z}$, so $g$ is at most the summation above.

We now show that $g$ is bounded below by this summation. Suppose we are given $0\le k'\le k$, a set $M\in\mathcal{I}_{k'}$, a graph $G_1\in\GTilde{t}{x}{k'}{z}{p}{M_X}{M}{M_Z}$, a graph $G_2\in\G{t-1}{x}{k-k'}{z}{p}{M_X}{M_K\setminus M}{M_Z}$, and a subset $\hat A\subseteq V_{args}\setminus(X_{args}\cup M_K)$ of size $k'-|M|$. Let $A = M\cup\hat A$, and let $B = V_{args}\setminus(X_{args}\cup A)$. We construct a graph $G$ as follows: Relabel $G_1$ by applying $\phi(X_{args_1}\setminus M_X,X_{args}\setminus M_X)$ to the labels in $X_{args_1}\setminus M_X$ and applying $\phi(V_{args_1}\setminus(X_{args_1}\cup M),\hat A)$ to the labels in $G_1\setminus(X_{args_1}\cup M)$. Relabel $G_2$ by applying $\phi(X_{args_2}\setminus M_X,X_{args}\setminus M_X)$ to the labels in $X_{args_2}\setminus M_X$ and applying $\phi(V_{args_2}\setminus(X_{args_2}\cup M_K),B\setminus M_K)$ to the labels in $G_2\setminus(X_{args_2}\cup M_K)$. Since $X_{args}$ is now a clique in both $G_1$ and $G_2$, we can glue $G_1$ and $G_2$ together at $X_{args}$ to obtain a chordal graph $G$. By \cref{lemma:supergraph_auto}, $\pi^p|_{V(G)}$ is an automorphism of $G$. By the proof of Lemma 3.6 in~\cite{hebertjohnson2023counting}, we have $G\in G(t,x,k,z)$.\footnote{See \cref{footnote:technically}.} Therefore, we have $G\in\G{t}{x}{k}{z}{p}{M_X}{M_K}{M_Z}$, so $g$ is at least the summation above.
\end{proof}

In the proof of the recurrence for $\tilde g$, we use the following relabeling procedures. The intuition behind these is the same as in our previous lemmas: we relabel the subgraphs in a similar way to~\cite{hebertjohnson2023counting}, but we do not relabel the moved vertices.

\medskip
\underline{\textit{Procedure $A$}:} Relabel $G_1$ by applying $\phi(\hat X',X_{args_1}\setminus M')$ to the labels in $\hat X'$ and applying $\phi(\hat C,V_{args_1}\setminus(X_{args_1}\cup M))$ to the labels in $\hat C$. Relabel $G_2$ by applying $\phi(X_{args}\setminus M_X,X_{args_2}\setminus M_X)$ to the labels in $X_{args}\setminus M_X$ and applying $\phi(D\setminus M_K,V_{args_2}\setminus(X_{args_2}\cup M_K))$ to the labels in $D\setminus M_K$.

\medskip
\underline{\textit{Procedure $B$}:} Relabel $G_1$ by applying $\phi(\hat X',X_{args_3}\setminus M')$ to the labels in $\hat X'$, and leave the labels in $C$ as they are. Relabel $G'$ by applying $\phi(X_{args}\setminus M_X,X_{args_4}\setminus M_X)$ to the labels in $X_{args}\setminus M_X$ and applying $\phi(D\setminus M_K,V_{args_4}\setminus(X_{args_4}\cup M_K))$ to the labels in $D\setminus M_K$.

\medskip
\underline{\textit{Procedure $A'$}:} In this procedure, we apply the inverse of all of the functions from \textit{Procedure $A$}. For example, when relabeling $G_1$, we apply $\phi(X_{args_1}\setminus M',\hat X')$ to the labels in $X_{args_1}\setminus M'$.

\medskip
\underline{\textit{Procedure $B'$}:} For all $i\in[p_C]$, relabel $G_i$ by applying $\phi(X_{args_3}\setminus M',\hat X')$ to the labels in $X_{args_3}\setminus M'$. Also, for all $2\le i\le p_C$, relabel $G_i$ by applying $\pi^{p(i-1)}$ to the labels in $M'$ and applying $\pi^{p(i-1)}$ to the labels in $C$. For $G'$, we apply the inverse of both of the functions that were applied to $G'$ in \textit{Procedure $B$}.

\gTildePi*

\begin{proof}
Let $\mathcal{G}_{inv}$ be the set of graphs $G\in\GTilde{t}{x}{k}{z}{p}{M_X}{M_K}{M_Z}$ such that $C$ is invariant under $\pi^p$, where $C$ is the connected component of $G\setminus X_{args}$ that contains $s$, and let $\mathcal{G}_{move} = \GTilde{t}{x}{k}{z}{p}{M_X}{M_K}{M_Z}\setminus\mathcal{G}_{inv}$. To prove this recurrence, we will first describe a map from $\mathcal{G}_{inv}$ to the set of tuples $(G_1,G_2,k',x',M,M',\hat C,\hat X')$ with the properties given below, and then we will describe a map from $\mathcal{G}_{move}$ to the set of tuples $(G_1,G',x',C,M',\hat X')$ given below. This will imply that $\tilde g$ is at most the summation above. Next, to show that $\tilde g$ is at least the summation above, we will describe a map from the set of tuples $(G_1,G_2,k',x',M,M',\hat C,\hat X')$ to $\mathcal{G}_{inv}$, as well as a map from the set of tuples $(G_1,G',x',C,M',\hat X')$ to $\mathcal{G}_{move}$. See \cref{lemma:injective_le,lemma:injective_ge} for the proofs that all of these maps are injective.

Let $args$ be the arguments of $\tilde g$ (on the left side), and let $args_1$, $args_2$, $args_3$, and $args_4$ be the arguments of $\tilde g_1$, $\tilde g$, $\tilde g_1$ (the lower one), and $\tilde g$ (the lower one), in the order that they appear on the right. To see that $\tilde g$ is at most the summation above, suppose $G\in\GTilde{t}{x}{k}{z}{p}{M_X}{M_K}{M_Z}$. Let $C$ be the set of vertices in the component of $G\setminus X_{args}$ that contains $s$, let $X' = N(C)$, and let $M' = X'\cap M_X$. There are two possible cases: either $C$ is invariant under $\pi^p$, or all of $C$ is mapped to some other component of $G\setminus X_{args}$ by $\pi^p$. If we are in the first case, then let $D$ be the set of vertices of all other components of $G\setminus X_{args}$. Let $G_1 = G[X'\cup C]$, $G_2 = G[X_{args}\cup D]$, $k' = |C|$, $x' = |N(C)|$, $M = C\cap M_K$, $\hat C = C\setminus M$, and $\hat X' = X'\setminus M'$. Note that $|\hat C| = k'-|M|$ and $|\hat X'| = x'-|M'|$. We know $\hat X'\not\subseteq Z_{args}$ if $M'\subseteq M_Z$ since the definition of $\tilde g$ tells us that $X'\not\subseteq Z_{args}$. Since $C$ is invariant under $\pi^p$, this means $N(C)$ is also invariant under $\pi^p$. Therefore, by \cref{lemma:subgraph_auto}, $\pi^p|_{V(G_1)}$ is an automorphism of $G_1$ and $\pi^p|_{V(G_2)}$ is an automorphism of $G_2$. Since $|M'|\le x'$, we have $M'\in\mathcal{I}_{x'}$. Clearly $|M|\le k'$, and we also have $k'-|M|\le k-|M_K|$ since the number of vertices that are fixed points of $\pi^p$ in $C$ is at most the number of fixed points in $G\setminus X_{args}$. If $s\in M_K$, then we know $s\in M$ by the definition of $C$. Otherwise, if $s\notin M_K$, then we have $|M|\le k'-1$ since $s\in C\setminus M$. Hence $M\in\mathcal{P}_{k'}$. Now relabel $G_1$ and $G_2$ using \textit{Procedure $A$} so that they have the desired vertex sets. By the proof of Lemma 3.7 in~\cite{hebertjohnson2023counting}, we have $G_1\in\widetilde G_1(t,x',k')$ and $G_2\in\widetilde G(t,x,k-k',z)$.\footnote{See \cref{footnote:technically} in \cref{lemma:g_pi}.} Therefore, we have $G_1\in\GTildeOne{t}{x'}{k'}{p}{M'}{M}$ and $G_2\in\GTilde{t}{x}{k-k'}{z}{p}{M_X}{M_K\setminus M}{M_Z}$.

If we are in the second case, where $C$ is mapped to some other component of $G\setminus X_{args}$ by $\pi^p$, then we claim that $(C,M')\in\mathcal{Q}$. We know every component of $G\setminus X_{args}$ is mapped to a component of $G\setminus X_{args}$ by $\pi^p$, so all vertices in $C$ have the same period $p_C\ge 2$ with respect to $(C,\pi^p)$. We have $N(\pi^{p\cdot p_C}(C)) = \pi^{p\cdot p_C}(N(C))$ since $\pi^{p\cdot p_C}|_{V(G)}$ is an automorphism of $G$, and we know $\pi^{p\cdot p_C}(C) = C$, so $\pi^{p\cdot p_C}(M') = M'$. Hence $(C,M')\in\mathcal{Q}$.

Let $x' = |N(C)|$ and let $\hat X' = X'\setminus M'$. Note that $|\hat X'| = x'-|M'|$. We also know $\hat X'\not\subseteq Z_{args}$ if $M'\subseteq M_Z$ since $X'\not\subseteq Z_{args}$. Let $C_1 = C$ and let $X_1' = X'$. For $2\le i\le p_C$, let $C_i = \pi^p(C_{i-1})$ and let $X_i' = \pi^p(X_{i-1}')$. Let $G_i = G[X_i'\cup C_i]$ for $i\in[p_C]$. For the remainder of the graph, let $D = V_{args}\setminus(X_{args}\cup C_1\cup\cdots\cup C_{p_C})$, and let $G' = G[X_{args}\cup D]$. Now relabel $G_1$ and $G'$ using \textit{Procedure $B$}.

Let $\hat G = G[\check X\cup\check C]$, where $\check X = X_1'\cup\cdots\cup X_{p_C}'$ and $\check C = C_1\cup\cdots\cup C_{p_C}$. By \cref{lemma:subgraph_auto}, since $\check X$ and $\check C$ are both invariant under $\pi^p$, we know $\pi^p|_{V(\hat G)}$ is an automorphism of $\hat G$ and $\pi^p|_{V(G')}$ is an automorphism of $G'$. Also, by applying \cref{lemma:subgraph_auto} to $\hat G$, we can see that $\pi^{p\cdot p_C}|_{V(G_1)}$ is an automorphism of $G_1$ since $\pi^{p\cdot p_C}|_{V(\hat G)}$ is an automorphism of $\hat G$ that maps $C$ to itself and maps $N(C)$ to itself. By the proof of Lemma 3.7 in~\cite{hebertjohnson2023counting}, we have $G_1\in\widetilde G_1(t,x',|C|)$ and $G'\in\widetilde G(t,x,k-p_C|C|,z)$.\footnote{See \cref{footnote:technically} in \cref{lemma:g_pi}.} Indeed, even though $G'$ is defined slightly differently from $G_2$, we can similarly show that $G'$ has the desired properties including the correct evaporation time. Therefore, we have $G_1\in\GTildeOne{t}{x'}{|C|}{p\cdot p_C}{M'}{C}$ and $G'\in\GTilde{t}{x}{k-p_C|C|}{z}{p}{M_X}{M_K\setminus C^{\pi}}{M_Z}$, so $\tilde g$ is at most the summation above.

We now show that $\tilde g$ is bounded below by this summation. For the first case, suppose we are given $1\le k'\le k$, $1\le x'\le x$, a set $M\in\mathcal{P}_{k'}$, a set $M'\in\mathcal{I}_{x'}$, a graph $G_1\in\GTildeOne{t}{x'}{k'}{p}{M'}{M}$, a graph $G_2\in\GTilde{t}{x}{k-k'}{z}{p}{M_X}{M_K\setminus M}{M_Z}$, a subset $\hat C\subseteq V_{args}\setminus(X_{args}\cup M_K)$ of size $k'-|M|$ that contains $s$ if $s\notin M_K$, and a subset $\hat X'\subseteq X_{args}\setminus M_X$ of size $x'-|M'|$ such that $M'\cup\hat X'\not\subseteq Z_{args}$. Let $C = M\cup\hat C$, $D = V_{args}\setminus(X_{args}\cup C)$, and $X' = M'\cup\hat X'$. We construct a graph $G$ as follows: Relabel $G_1$ and $G_2$ using \textit{Procedure $A'$}, and then glue $G_1$ and $G_2$ together at $X'$ to obtain $G$. By \cref{lemma:supergraph_auto}, $\pi^p|_{V(G)}$ is an automorphism of $G$. By the proof of Lemma 3.7 in~\cite{hebertjohnson2023counting}, we have $G\in\widetilde G(t,x,k,z)$.\footnote{See \cref{footnote:technically} in \cref{lemma:g_pi}.} Therefore, we have $G\in\GTilde{t}{x}{k}{z}{p}{M_X}{M_K}{M_Z}$.

For the second case, suppose we are given $1\le x'\le x$, a pair $(C,M')\in\mathcal{Q}$, a graph $G_1\in\GTildeOne{t}{x'}{|C|}{p\cdot p_C}{M'}{C}$, a graph $G'\in\GTilde{t}{x}{k-p_C|C|}{z}{p}{M_X}{M_K\setminus C^{\pi}}{M_Z}$, and a subset $\hat X'\subseteq X_{args}\setminus M_X$ of size $x'-|M'|$ such that $M'\cup\hat X'\not\subseteq Z_{args}$. Let $X' = M'\cup\hat X'$, and let $D = V_{args}\setminus(X_{args}\cup C_{\pi^p})$. We construct a graph $G$ as follows: Let $G_2,\ldots,G_{p_C}$ all be equal to $G_1$ initially. Next, relabel $G_1,\ldots,G_{p_C}$ and $G'$ using \textit{Procedure $B'$}. For all $i\in[p_C]$, glue the graph $G_i$ to $G'$ at $\pi^{p(i-1)}(X')$, and let $G$ be the resulting graph. We know $X'\not\subseteq Z_{args}$, which also implies $\pi^{p(i-1)}(X')\not\subseteq Z_{args}$ for all $i\in[p_C]$ since $M_Z$ is invariant under $\pi^p$.

An alternative way to construct the same graph $G$ would be to do the following: Let $\check X$ be a clique with vertex set $X'\cup\pi^p(X')\cup\cdots\cup\pi^{p(p_C-1)}(X')$. Let $\hat G$ be the graph obtained by gluing $G_i$ to $\check X$ at the clique $\pi^{p(i-1)}(X')$ for all $i\in[p_C]$. To obtain $G$, glue $\hat G$ and $G'$ together at $\check X$. We claim that $\pi^p|_{V(\hat G)}$ is an automorphim of $\hat G$. We know $\pi^p(V(\hat G)) = V(\hat G)$ since $\pi^{p\cdot p_C}(C) = C$ and $\pi^{p\cdot p_C}(X') = X'$. Let $u,v\in V(\hat G)$. If $u,v\in\check X$, then $\pi^p(u),\pi^p(v)\in\check X$, so $u$ and $v$ are adjacent, and so are $\pi^p(u)$ and $\pi^p(v)$. If $u\in\pi^{pi}(C)$ and $v\in\pi^{pj}(C)$ for some $i\not\equiv j\pmod{p_C}$, then $\pi^p(u)\in\pi^{p(i+1)}(C)$ and $\pi^p(v)\in\pi^{p(j+1)}(C)$, so $u$ and $v$ are not adjacent, and neither are $\pi^p(u)$ and $\pi^p(v)$. If $u,v\in V(G_i)$ for some $1\le i<p_C$, then $u,v\in V(G_{i+1})$, so $u$ and $v$ are adjacent if and only if $\pi^p(u)$ and $\pi^p(v)$ are adjacent by the way $G_{i+1}$ was constructed. If $u,v\in V(G_{p_C})$, then $u,v\in V(G_1)$. This means $u$ and $v$ are adjacent if and only if $\pi^p(u)$ and $\pi^p(v)$ are adjacent since $\pi^{p\cdot p_C}|_{V(G_1)}$ is an automorphism of $G_1$. Lastly, if $u\in\pi^{pi}(C)$ for some $i$ and $v\in\check X\setminus\pi^{pi}(X')$ (or vice versa), then $u$ and $v$ are not adjacent, and neither are $\pi^p(u)$ and $\pi^p(v)$ since $\pi^{p\cdot p_C}|_{V(G_1)}$ is an automorphism of $G_1$. Hence $\pi^p|_{V(\hat G)}$ is an automorphim of $\hat G$. By \cref{lemma:supergraph_auto}, this implies that $\pi^p|_{V(G)}$ is an automorphism of $G$. By an argument similar to Lemma 3.7 in~\cite{hebertjohnson2023counting}, we have $G\in\widetilde G(t,x,k,z)$.\footnote{See \cref{footnote:technically} in \cref{lemma:g_pi}.} Therefore, we have $G\in\GTilde{t}{x}{k}{z}{p}{M_X}{M_K}{M_Z}$, so $\tilde g$ is at least the summation above.
\end{proof}

In the following two lemmas, let $\mathcal{G}_{inv}$ and $\mathcal{G}_{move}$ be as defined in \cref{lemma:g_tilde_pi}, and let $args = \begin{pmatrix}
t\, & x & k & z \\
p\, & M_X & M_K & M_Z
\end{pmatrix}$ be fixed values of the arguments of $\tilde g$.

\begin{lemma}
\label{lemma:injective_le}
Let $\psi_1$ be the map from $\mathcal{G}_{inv}$ to the set of tuples $(G_1,G_2,k',x',M,M',\hat C,\hat X')$ with certain properties from the first case of the ``$\le$'' direction of the proof of \cref{lemma:g_tilde_pi}, and let $\psi_2$ be the map from $\mathcal{G}_{move}$ to the set of tuples $(G_1,G',x',C,M',\hat X')$ with certain properties from the second case. The maps $\psi_1$ and $\psi_2$ are injective.
\end{lemma}

\begin{proof} To see that $\psi_1$ is injective, suppose $G$ and $G^\star$ are distinct graphs in $\mathcal{G}_{inv}$. We want to show $\psi_1(G)\ne\psi_1(G^\star)$. Suppose that starting from $G$, we obtain $\psi_1(G) = (G_1,G_2,k',x',M,M',\hat C,\hat X')$, along with the sets $C$, $D$, and $X'$ as described in \cref{lemma:g_tilde_pi}. Also, suppose that starting from $G^\star$, we obtain $\psi_1(G^\star) = (G_1^\star,G_2^\star,k'^\star,x'^\star,M^\star,M'^\star,\hat C^\star,\hat X'^\star)$, along with the sets $C^\star$, $D^\star$, and $X'^\star$. If $(k',x',M,M',\hat C,\hat X')\ne(k'^\star,x'^\star,M^\star,M'^\star,\hat C^\star,\hat X'^\star)$, then we are done, so suppose $(k',x',M,M',\hat C,\hat X') = (k'^\star,x'^\star,M^\star,M'^\star,\hat C^\star,\hat X'^\star)$. This means $C = C^\star$ since $C = M\cup\hat C$ and $C^\star = M^\star\cup\hat C^\star$, which implies $D = D^\star$. Similarly, we also have $X' = X'^\star$.

Since $G\ne G^\star$, we either have $G[X'\cup C]\ne G^\star[X'\cup C]$ or $G[X_{args}\cup D]\ne G^\star[X_{args}\cup D]$. If $G[X'\cup C]\ne G^\star[X'\cup C]$, then after relabeling, $G[X'\cup C]$ still differs from $G^\star[X'\cup C]$ since we use the same sets $M = M^\star$ and $M' = M'^\star$ in both of these relabeling procedures. Thus in this case, we have $G_1\ne G_1^\star$. Otherwise, if $G[X_{args}\cup D]\ne G^\star[X_{args}\cup D]$, then after relabeling, $G[X_{args}\cup D]$ still differs from $G^\star[X_{args}\cup D]$ since we use the same sets $M_X$ and $M_K$ when relabeling, so we have $G_2\ne G_2^\star$. Therefore, $\psi_1$ is injective.

To see that $\psi_2$ is injective, suppose $G$ and $G^\star$ are distinct graphs in $\mathcal{G}_{move}$. We want to show $\psi_2(G)\ne\psi_2(G^\star)$. Suppose that starting from $G$, we obtain $\psi_2(G) = (G_1,G',x',C,M',\hat X')$, along with the sets $C_1,\ldots,C_{p_C},D,X_1',\ldots,X_{p_C}'$ and the graphs $G_2,\ldots,G_{p_C}$ as described in \cref{lemma:g_tilde_pi}, where $C_1 = C$. Also, suppose that starting from $G^\star$, we obtain $\psi_2(G^\star) = (G_1^\star,G'^\star,x'^\star,C^\star,M'^\star,\hat X'^\star)$, along with the sets $C_1^\star,\ldots,C_{p_C}^\star,D^\star,X_1'^\star,\ldots,X_{p_C}'^\star$, and the graphs $G_2^\star,\ldots,G_{p_C}^\star$, where $C_1^\star = C^\star$. If $(x',C,M',\hat X')\ne(x'^\star,C^\star,M'^\star,\hat X'^\star)$, then we are done, so suppose $(x',C,M',\hat X') = (x'^\star,C^\star,M'^\star,\hat X'^\star)$. We have $C_i = C_i^\star$ for all $i\in[p_C]$ since $C = C^\star$, using the fact that $C_i = \pi^p(C_{i-1})$ and $C_i^\star = \pi^p(C_{i-1}^\star)$. This implies $D = D^\star$. Similarly, we also have $X_i' = X_i'^\star$ for all $i\in[p_C]$ since $X_1' = X_1'^\star$.

Since $G\ne G^\star$, we either have $G[X_i'\cup C_i]\ne G^\star[X_i'\cup C_i]$ for some $i\in[p_C]$ or $G[X_{args}\cup D]\ne G^\star[X_{args}\cup D]$. First, suppose we have $G[X_i'\cup C_i]\ne G^\star[X_i'\cup C_i]$ for some $i\in[p_C]$. For all $2\le i\le p_C$, $\pi^{p(i-1)}$ is an isomorphism from $G_1$ to $G_i$, since $\pi^p,\ldots,\pi^{p(p_C-1)}$ are automorphisms of $G$. Similarly, for all $2\le i\le p_C$, $\pi^{p(i-1)}$ is an isomorphism from $G_1^\star$ to $G_i^\star$. Therefore, we have $G[X_1'\cup C_1]\ne G^\star[X_1'\cup C_1]$. After relabeling, $G[X_1'\cup C_1]$ still differs from $G^\star[X_1'\cup C_1]$ since we use the same set $M' = M'^\star$ in both of these relabeling procedures. Thus in this case, we have $G_1\ne G_1^\star$. Otherwise, if $G[X_{args}\cup D]\ne G^\star[X_{args}\cup D]$, then after relabeling, $G[X_{args}\cup D]$ still differs from $G^\star[X_{args}\cup D]$ since we use the same sets $M_X$ and $M_K$ when relabeling, so we have $G_2\ne G_2^\star$. Therefore, $\psi_2$ is injective.
\end{proof}

\begin{lemma}
\label{lemma:injective_ge}
Let $\psi_1'$ be the map from the set of tuples $(G_1,G_2,k',x',M,M',\hat C,\hat X')$ with certain properties to $\mathcal{G}_{inv}$ from the first case of the ``$\ge$'' direction of the proof of \cref{lemma:g_tilde_pi}, and let $\psi_2'$ be the map from the set of tuples $(G_1,G',x',C,M',\hat X')$ with certain properties to $\mathcal{G}_{move}$ from the second case. The maps $\psi_1'$ and $\psi_2'$ are injective.
\end{lemma}

\begin{proof}
To see that $\psi_1'$ is injective, suppose $$(G_1,G_2,k',x',M,M',\hat C,\hat X')\text{ and }(G_1^\star,G_2^\star,k'^\star,x'^\star,M^\star,M'^\star,\hat C^\star,\hat X'^\star)$$ are distinct tuples in the domain of $\psi_1'$ that map to $G$ and $G^\star$, respectively. If $k'\ne k'^\star$, then either $M\ne M^\star$ or $\hat C\ne\hat C^\star$ since $k' = |M|+|\hat C|$ and $k'^\star = |M^\star|+|\hat C^\star|$. Similarly, if $x'\ne x'^\star$, then $\hat X'\ne\hat X'^\star$ or $M'\ne M'^\star$. Therefore, we know $(G_1,G_2,M,M',\hat C,\hat X')\ne(G_1^\star,G_2^\star,M^\star,M'^\star,\hat C^\star,\hat X'^\star)$. Let $X' = M'\cup\hat X'$ and $X'^\star = M'^\star\cup\hat X'^\star$.

First, suppose $(M,\hat C)\ne(M^\star,\hat C^\star)$, which means $M\cup\hat C\ne M^\star\cup\hat C^\star$. After we relabel $G_1$, the vertex set of $G_1\setminus X'$ is $M\cup\hat C$, which contains $s$. Similarly, after we relabel $G_1^\star$, the vertex set of $G_1^\star\setminus X'^\star$ is $M^\star\cup\hat C^\star$, which also contains $s$. Therefore, the component of $G\setminus X_{args}$ that contains $s$ has a different label set from the component of $G^\star\setminus X_{args}$ that contains $s$, so $G\ne G^\star$.

Next, suppose $(M',\hat X')\ne(M'^\star,\hat X'^\star)$, which means $X'\ne X'^\star$. The neighborhood of the component of $G\setminus X_{args}$ that contains $s$ is $X'$, and the neighborhood of the component of $G^\star\setminus X_{args}$ that contains $s$ is $X'^\star$. Hence $G\ne G^\star$.

Lastly, suppose $(G_1,G_2)\ne(G_1^\star,G_2^\star)$ and $(M,M',\hat C,\hat X') = (M^\star,M'^\star,\hat C^\star,\hat X'^\star)$, since this is the only remaining possibility. If $G_1\ne G_1^\star$, then either the component of $G\setminus X_{args}$ that contains $s$ will differ from the component of $G^\star\setminus X_{args}$ that contains $s$, or the edges from those components to $X_{args}$ will differ. Similarly, if $G_2\ne G_2^\star$, then either the remaining components of $G\setminus X_{args}$ will differ from those of $G^\star\setminus X_{args}$, or the edges from those components to $X_{args}$ will differ. Hence $G\ne G^\star$, so $\psi_1'$ is injective.

To see that $\psi_2'$ is injective, suppose $(G_1,G',x',C,M',\hat X')$ and $(G_1^\star,G'^\star,x'^\star,C^\star,M'^\star,\hat X'^\star)$ are distinct tuples in the domain of $\psi_2'$ that map to $G$ and $G^\star$, respectively. By a similar argument to the first paragraph above, we know $(G_1,G',C,M',\hat X')\ne(G_1^\star,G'^\star,C^\star,M'^\star,\hat X'^\star)$. As before, let $X' = M'\cup\hat X'$ and $X'^\star = M'^\star\cup\hat X'^\star$.

If $C\ne C^\star$, then a similar argument to the second paragraph shows that $G\ne G^\star$. If $(M',\hat X')\ne(M'^\star,\hat X'^\star)$, then a similar argument to the third paragraph shows that $G\ne G^\star$. Lastly, if $(G_1,G')\ne(G_1^\star,G'^\star)$ and $(C,M',\hat X')\ne(C^\star,M'^\star,\hat X'^\star)$, then a similar argument to the fourth paragraph shows that $G\ne G^\star$. Hence $\psi_2'$ is injective.
\end{proof}

\gTildePPi*

\begin{proof}
The argument for $\tilde g_p$ is similar to the proof for $\tilde g$, except we do not allow $x' = x$ since no component of $G\setminus X_{args}$ sees all of $X_{args}$.
\end{proof}

\gTildeOnePi*

\begin{proof}
Let $\mathcal{G} = \GTildeOne{t}{x}{k}{p}{M_X}{M_K}$. To prove this recurrence, we will first describe a map from $\mathcal{G}$ to the set of tuples $(G,l,M,\hat L)$ with the properties given below, which will imply that $\tilde g_1$ is at most the summation above. Next, to show that $\tilde g_1$ is at least the summation above, we will describe a map from the set of tuples $(G,l,M,\hat L)$ to $\mathcal{G}$. Strictly speaking, to prove these two bounds ($\le$ and $\ge$), it remains to show that both of these maps are injective. See \cref{remark:injective}.

Let $args$ be the arguments of $\tilde g_1$. To see that $\tilde g_1$ is at most the summation above, suppose $G\in\GTildeOne{t}{x}{k}{p}{M_X}{M_K}$. Let $C = V_{args}\setminus X_{args}$, $L = L_G(X_{args})$, $A = C\setminus L$, $l = |L|$, $M = L\cap M_K$, and $\hat L = L\setminus M$. Note that $|\hat L| = l - |M|$. We know $\pi^p|_{V(G)}$ is an automorphism of $G$. By \cref{lemma:invariant_evap_time}, $L$ is invariant under $\pi^p$, so $M\subseteq_{\pi^p}M_K$. Clearly $|M'|\le l$, and we also have $l-|M|\le k-|M_K|$ since the number of vertices that are fixed points of $\pi^p$ in $L$ is at most the number of fixed points in $G\setminus X_{args}$. Hence $M\in\mathcal{I}_l$. Now relabel $G$ to have the desired vertex set, in a similar way to \cref{lemma:g_pi}. By the proof of Lemma 3.4 in~\cite{hebertjohnson2023counting}, we have $G\in F(t,x,l,k-l)$.\footnote{See \cref{footnote:technically} in \cref{lemma:g_pi}.} Therefore, $G\in\F{t}{x}{l}{k-l}{p}{M_X}{M}{M_K\setminus M}$, so $\tilde g_1$ is at most the summation above.

We now show that $\tilde g_1$ is bounded below by this summation. Suppose we are given $1\le l\le k$, a set $M\in\mathcal{I}_l$, a graph $G\in\F{t}{x}{l}{k-l}{p}{M_X}{M}{M_K\setminus M}$, and a subset $\hat L\subseteq V_{args}\setminus(X_{args}\cup M_K)$ of size $l-|M|$. Relabel $G$ to have the desired vertex set (in a similar way to \cref{lemma:g_pi}). Clearly $\pi^p|_{V(G)}$ is an automorphism of $G$. By the proof of Lemma 3.4 in~\cite{hebertjohnson2023counting}, we have $G\in\widetilde G_1(t,x,k)$.\footnote{See \cref{footnote:technically} in \cref{lemma:g_pi}.} Therefore, $G\in\GTildeOne{t}{x}{k}{p}{M_X}{M_K}$, so $\tilde g_1$ is at least the summation above.
\end{proof}

\gTildeTwoPi*

\begin{proof}
Let $\mathcal{G}_{inv}$ be the set of graphs $G\in\GTildeTwo{t}{x}{k}{p}{M_X}{M_K}$ such that $C$ is invariant under $\pi^p$, where $C$ is the connected component of $G\setminus X_{args}$ that contains $s$, and let $\mathcal{G}_{move} = \GTildeTwo{t}{x}{k}{p}{M_X}{M_K}\setminus\mathcal{G}_{inv}$. To prove this recurrence, we will first describe a map from $\mathcal{G}_{inv}$ to the set of tuples $(G_1,G_2,k',M,\hat C)$ with the properties given below, and then we will describe a map from $\mathcal{G}_{move}$ to the set of tuples $(G_1,G',C)$ given below. This will imply that $\tilde g_{\ge 2}$ is at most the summation above. Next, to show that $\tilde g_{\ge 2}$ is at least the summation above, we will describe a map from the set of tuples $(G_1,G_2,k',M,\hat C)$ to $\mathcal{G}_{inv}$, as well as a map from the set of tuples $(G_1,G',C)$ to $\mathcal{G}_{move}$. Strictly speaking, to prove these two bounds ($\le$ and $\ge$), it remains to show that all of these maps are injective. See \cref{remark:injective}.

Let $args$ be the arguments of $\tilde g_{\ge 2}$. To see that $\tilde g_{\ge 2}$ is at most the summation above, suppose $G\in\GTildeTwo{t}{x}{k}{p}{M_X}{M_K}$. Let $C$ be the set of vertices in the component of $G\setminus X_{args}$ that contains $s$. There are two possible cases: either $C$ is invariant under $\pi^p$, or all of $C$ is mapped to some other component of $G\setminus X_{args}$ by $\pi^p$. If we are in the first case, then let $D$ be the set of vertices of all other components of $G\setminus X_{args}$. Let $G_1 = G[X_{args}\cup C]$, $G_2 = G[X_{args}\cup D]$, $k' = |C|$, $M = C\cap M_K$, and $\hat C = C\setminus M$. Note that $|\hat C| = k'-|M|$. We know $X_{args}$ is invariant under $\pi^p$, so by \cref{lemma:subgraph_auto}, $\pi^p|_{V(G_1)}$ is an automorphism of $G_1$ and $\pi^p|_{V(G_2)}$ is an automorphism of $G_2$. We also have $M\in\mathcal{P}_{k'}$ by an argument similar to the proof for $\tilde g$. Now relabel $G_1$ and $G_2$ in a similar way to \cref{lemma:g_tilde_pi}. By the proof of Lemma 3.9 in~\cite{hebertjohnson2023counting}, we have $G_2\in\widetilde G_1(t,x,k-k')$ if $G\setminus X_{args}$ has exactly two connected components, and we have $G_2\in\widetilde G_{\ge 2}(t,x,k-k')$ otherwise. In either case, we have $G_1\in\widetilde G_1(t,x,k')$.\footnote{See \cref{footnote:technically} in \cref{lemma:g_pi}.} Therefore, we have $G_1\in\GTildeOne{t}{x}{k'}{p}{M_X}{M}$ and either $G_2\in\GTildeOne{t}{x}{k-k'}{p}{M_X}{M_K\setminus M}$ or $G_2\in\GTildeTwo{t}{x}{k-k'}{p}{M_X}{M_K\setminus M}$.

If we are in the second case, where $C$ is mapped to some other component of $G\setminus X_{args}$ by $\pi^p$, then we know $C\in\mathcal{Q}$ by an argument similar to the proof for $\tilde g$. Let $C_1 = C$, and let $C_i = \pi^p(C_{i-1})$ for $2\le i\le p_C$. Let $G_i = G[X_{args}\cup C_i]$ for $i\in[p_C]$. For the remainder of the graph, let $D = V_{args}\setminus(X_{args}\cup C_1\cup\cdots\cup C_{p_C})$, and let $G' = G[X_{args}\cup D]$. Now relabel $G_1$ and $G'$ in a similar way to \cref{lemma:g_tilde_pi}. We can see that $\pi^{p\cdot p_C}|_{V(G_1)}$ is an automorphism of $G_1$ and $\pi^p|_{V(G')}$ is an automorphism of $G'$ by an argument similar to the proof for $\tilde g$. By an argument similar to Lemma 3.9 in~\cite{hebertjohnson2023counting}, we have $G'\in\widetilde G_1(t,x,k-p_C|C|)$ if $G\setminus X_{args}$ has exactly two connected components, and we have $G'\in\widetilde G_{\ge 2}(t,x,k-p_C|C|)$ otherwise. In either case, we have $G_1\in\widetilde G_1(t,x,|C|)$.\footnote{See \cref{footnote:technically} in \cref{lemma:g_pi}.} Therefore, we have $G_1\in\GTildeOne{t}{x}{|C|}{p\cdot p_C}{M_X}{C}$ and either $G_2\in\GTildeOne{t}{x}{k-p_C|C|}{p}{M_X}{M_K\setminus C^{\pi}}$ or $G_2\in\GTildeTwo{t}{x}{k-p_C|C|}{p}{M_X}{M_K\setminus C^{\pi}}$, so $\tilde g_{\ge 2}$ is at most the summation above.

We now show that $\tilde g_{\ge 2}$ is bounded below by this summation. For the first case, suppose we are given $1\le k'\le k$, a set $M\in\mathcal{P}_{k'}$, a graph $G_1\in\GTildeOne{t}{x}{k'}{p}{M_X}{M}$, a graph $G_2$ that belongs to $\GTildeOne{t}{x}{k-k'}{p}{M_X}{M_K\setminus M}$ or $\GTildeTwo{t}{x}{k-k'}{p}{M_X}{M_K\setminus M}$, and a subset $\hat C\subseteq V_{args}\setminus(X_{args}\cup M_K)$ of size $k'-|M|$ that contains $s$ if $s\notin M_K$. We construct a graph $G$ as follows: Relabel $G_1$ and $G_2$ in a similar way to \cref{lemma:g_tilde_pi}, and then glue $G_1$ and $G_2$ together at $X_{args}$ to obtain $G$. By \cref{lemma:supergraph_auto}, $\pi^p|_{V(G)}$ is an automorphism of $G$. By the proof of Lemma 3.9 in~\cite{hebertjohnson2023counting}, we have $G\in\widetilde G_{\ge 2}(t,x,k)$.\footnote{See \cref{footnote:technically} in \cref{lemma:g_pi}.} Therefore, we have $G\in\GTildeTwo{t}{x}{k}{p}{M_X}{M_K}$.

For the second case, suppose we are given a set $C\in\mathcal{Q}$, a graph $G_1\in\GTildeOne{t}{x}{|C|}{p\cdot p_C}{M_X}{C}$, and a graph $G'$ that belongs to $\GTildeOne{t}{x}{k-p_C|C|}{p}{M_X}{M_K\setminus C^{\pi}}$ or $\GTildeTwo{t}{x}{k-p_C|C|}{p}{M_X}{M_K\setminus C^{\pi}}$. We construct a graph $G$ as follows: Let $G_2,\ldots,G_{p_C}$ all be equal to $G_1$ initially. Next, relabel $G_1,\ldots,G_{p_C}$ and $G'$ in a similar way to \cref{lemma:g_tilde_pi}. For all $i\in[p_C]$, glue the graph $G_i$ to $G'$ at $X_{args}$, and let $G$ be the resulting graph. We can see that $\pi^p|_{V(G)}$ is an automorphism of $G$ by an argument similar to the proof for $\tilde g$. By an argument similar to Lemma 3.9 in~\cite{hebertjohnson2023counting}, we have $G\in\widetilde G_{\ge 2}(t,x,k)$.\footnote{See \cref{footnote:technically} in \cref{lemma:g_pi}.} Therefore, we have $G\in\GTildeTwo{t}{x}{k}{p}{M_X}{M_K}$, so $\tilde g$ is at least the summation above.
\end{proof}

\fPi*

\begin{proof}
Let $\mathcal{G} = \F{t}{x}{l}{k}{p}{M_X}{M_L}{M_K}$. To prove this recurrence, we will first describe a map from $\mathcal{G}$ to the set of tuples $(G_1,G_2,k',M,\hat A)$ with the properties given below, which will imply that $f$ is at most the summation above. Next, to show that $f$ is at least the summation above, we will describe a map from the set of tuples $(G_1,G_2,k',M,\hat A)$ to $\mathcal{G}$. Strictly speaking, to prove these two bounds ($\le$ and $\ge$), it remains to show that both of these maps are injective. See \cref{remark:injective}.

Let $args$, $args_1$, and $args_2$ be the arguments of $f$, $\tilde f$, and $g$, respectively. To see that $f$ is at most the summation above, suppose $G\in\F{t}{x}{l}{k}{p}{M_X}{M_L}{M_K}$. Let $A$ be the set of vertices in components $C$ of $G\setminus(X_{args}\cup L_{args})$ such that $C$ evaporates at time exactly $t-1$ in $G$ with exception set $X_{args}$, and let $B$ be the set of vertices of all other components of $G\setminus(X_{args}\cup L_{args})$. Let $G_1 = G[X_{args}\cup L_{args}\cup A]$, $G_2 = G[X_{args}\cup L_{args}\cup B]$, $k' = |A|$, $M$ = $A\cap M_K$, and $\hat A = A\setminus M$. Note that $|\hat A| = k' - |M|$. By \cref{lemma:invariant_evap_time}, $A$ is invariant under $\pi^p$, and so is $M$. Since $G$ has the automorphism $\pi^p|_{V(G)}$, by \cref{lemma:subgraph_auto} this means $G_1$ has the automorphism $\pi^p|_{V(G_1)}$ and $G_2$ has the automorphism $\pi^p|_{V(G_2)}$. We also have $M\in\mathcal{I}_{k'}$. Now relabel $G_1$ and $G_2$ so that they have the desired vertex sets, in a similar way to \cref{lemma:g_pi}. By the proof of Lemma 3.5 in~\cite{hebertjohnson2023counting}, we have $G_1\in\widetilde F(t,x,l,k')$ and $G_2\in G(t-2,x+l,k-k',x)$.\footnote{See \cref{footnote:technically} in \cref{lemma:g_pi}.} Therefore, we have $G_1\in\FTilde{t}{x}{l}{k'}{p}{M_X}{M_L}{M}$ and $G_2\in\G{t-2}{x+l}{k-k'}{x}{p}{M_X\cup M_L}{M_K\setminus M}{M_X}$, so $f$ is at most the summation above.

We now show that $f$ is bounded below by this summation. Suppose we are given $1\le k'\le k$, a set $M\in\mathcal{I}_{k'}$, a graph $G_1\in\FTilde{t}{x}{l}{k'}{p}{M_X}{M_L}{M}$, a graph $G_2\in\G{t-2}{x+l}{k-k'}{x}{p}{M_X\cup M_L}{M_K\setminus M}{M_X}$, and a subset $\hat A\subseteq V_{args}\setminus(X_{args}\cup L_{args}\cup M_K)$ of size $k'-|M|$. We construct a graph $G$ as follows: Relabel $G_1$ and $G_2$ so that they have the desired vertex sets, including the clique $X_{args}\cup L_{args}$ (in a similar way to \cref{lemma:g_pi}), and then glue $G_1$ and $G_2$ together at $X_{args}\cup L_{args}$ to obtain $G$. By \cref{lemma:supergraph_auto}, $\pi^p|_{V(G)}$ is an automorphism of $G$. By the proof of Lemma 3.5 in~\cite{hebertjohnson2023counting}, we have $G\in F(t,x,l,k)$.\footnote{See \cref{footnote:technically} in \cref{lemma:g_pi}.} Therefore, $G\in\F{t}{x}{l}{k}{p}{M_X}{M_L}{M_K}$, so $f$ is at least the summation above.
\end{proof}

\fTildePi*

\begin{proof}
Let $\mathcal{G}$ be the set of graphs in $\FTilde{t}{x}{l}{k}{p}{M_X}{M_L}{M_K}$ such that at least one component of $G\setminus(X_{args}\cup L_{args})$ sees all of $X_{args}\cup L_{args}$. To prove (the second and third cases of) this recurrence, we will first describe a map from $\mathcal{G}$ to the set of tuples $(G_1,G_2,k',M,\hat A)$ with the properties given below, which will imply that $f$ is at most the summation above. Next, to show that $f$ is at least the summation above, we will describe a map from the set of tuples $(G_1,G_2,k',M,\hat A)$ to $\mathcal{G}$. Strictly speaking, to prove these two bounds ($\le$ and $\ge$), it remains to show that both of these maps are injective. See \cref{remark:injective}.

Let $args$ be the arguments of $\tilde f$, and let $args_1$, $args_2$, $args_3$, $args_4$, and $args_5$ be the arguments of $\tilde f_p$ (the first one), $\tilde g_1$, $\tilde f_p$ (the second one), $\tilde g_{\ge 2}$, and $\tilde g_p$, respectively. To see that $\tilde f$ is at most the summation above, suppose $G\in\FTilde{t}{x}{l}{k}{p}{M_X}{M_L}{M_K}$. The graph $G$ falls into one of three cases: either zero, one, or at least two components of $G\setminus(X_{args}\cup L_{args})$ see all of $X_{args}\cup L_{args}$. In the first case, we have $G\in\FTildeP{t}{x}{l}{k}{p}{M_X}{M_L}{M_K}$. If $G$ falls into the second or third case, then let $A$ be the set of vertices in the components that see all of $X_{args}\cup L_{args}$, and let $B$ be the set of vertices of all other components of $G\setminus(X_{args}\cup L_{args})$. Let $G_1 = G[X_{args}\cup L_{args}\cup A]$, $G_2 = G[X_{args}\cup L_{args}\cup B]$, $k' = |A|$, $M$ = $A\cap M_K$, and $\hat A = A\setminus M$. Note that $|\hat A| = k' - |M|$. By \cref{lemma:invariant_evap_time}, $L_{args}$ is invariant under $\pi^p$, which implies that $A$ is invariant under $\pi^p$ (and so is $M$). Since $G$ has the automorphism $\pi^p|_{V(G)}$, by \cref{lemma:subgraph_auto} this means $G_1$ has the automorphism $\pi^p|_{V(G_1)}$ and $G_2$ has the automorphism $\pi^p|_{V(G_2)}$. We also have $M\in\mathcal{I}_{k'}$. Now relabel $G_1$ and $G_2$ so that they have the desired vertex sets, in a similar way to \cref{lemma:g_pi}. If $A$ consists of exactly one component of $G\setminus(X_{args}\cup L_{args})$, then by the proof of Lemma 3.8 in~\cite{hebertjohnson2023counting}, we have $G_1\in\widetilde G_1(t-1,x+l,k')$ and $G_2\in\widetilde F_p(t,x,l,k-k')$,\footnote{See \cref{footnote:technically} in \cref{lemma:g_pi}.} which implies $G_1\in\GTildeOne{t-1}{x+l}{k'}{p}{M_X\cup M_L}{M}$ and $G_2\in\FTildeP{t}{x}{l}{k-k'}{p}{M_X}{M_L}{M_K\setminus M}$. Otherwise, by the proof of Lemma 3.8 in~\cite{hebertjohnson2023counting}, we have $G_1\in\widetilde G_{\ge 2}(t-1,x+l,k')$ and $G_2\in\widetilde G_p(t-1,x+l,k-k',x)$,\footnote{See \cref{footnote:technically} in \cref{lemma:g_pi}.} which implies $G_1\in\GTildeTwo{t-1}{x+l}{k'}{p}{M_X\cup M_L}{M}$ and $G_2\in\GTildeP{t-1}{x+l}{k-k'}{x}{p}{M_X\cup M_L}{M_K\setminus M}{M_X}$. Hence $\tilde f$ is at most the summation above.

We now show that $\tilde f$ is bounded below by this summation. First of all, every graph in $\FTildeP{t}{x}{l}{k}{p}{M_X}{M_L}{M_K}$ belongs to $\FTilde{t}{x}{l}{k}{p}{M_X}{M_L}{M_K}$. For the second and third case, suppose we are given $1\le k'\le k$, a set $M\in\mathcal{I}_{k'}$, a subset $\hat A\subseteq V_{args}\setminus(X_{args}\cup L_{args}\cup M_K)$ of size $k'-|M|$, and either a pair of graphs $G_1\in\GTildeOne{t-1}{x+l}{k'}{p}{M_X\cup M_L}{M}$ and $G_2\in\FTildeP{t}{x}{l}{k-k'}{p}{M_X}{M_L}{M_K\setminus M}$ or a pair of graphs $G_1\in\GTildeTwo{t-1}{x+l}{k'}{p}{M_X\cup M_L}{M}$ and $G_2\in\GTildeP{t-1}{x+l}{k-k'}{x}{p}{M_X\cup M_L}{M_K\setminus M}{M_X}$. We construct a graph $G$ as follows: Relabel $G_1$ and $G_2$ so that they have the desired vertex sets, including the clique $X_{args}\cup L_{args}$ (in a similar way to \cref{lemma:g_pi}), and then glue $G_1$ and $G_2$ together at $X_{args}\cup L_{args}$ to obtain $G$. By \cref{lemma:supergraph_auto}, we know $\pi^p|_{V(G)}$ is an automorphism of $G$. By the proof of Lemma 3.8 in~\cite{hebertjohnson2023counting}, we have $G\in\widetilde F(t,x,l,k)$.\footnote{See \cref{footnote:technically} in \cref{lemma:g_pi}.} Therefore, $G\in\FTilde{t}{x}{l}{k}{p}{M_X}{M_L}{M_K}$, so $\tilde f$ is at least the summation above.
\end{proof}

\fTildePPi*

\begin{proof}
This is similar to the proof that $\tilde f_p(t,x,l,k) = \tilde f_p(t,x,l,k,x)$ in~\cite{hebertjohnson2023counting} (see Lemma 3.11 in~\cite{hebertjohnson2023counting}). The only difference is that both sides now have the automorphism $\pi^p|_{V(G)}$, and the relevant vertex subsets are $V_{args}$, $X_{args}$, and $L_{args}$.
\end{proof}

\fTildePWithZPi*

\begin{proof}
Let $\mathcal{G}_{inv}$ be the set of graphs $G\in\FTildePWithZ{t}{x}{l}{k}{z}{p}{M_X}{M_L}{M_K}{M_Z}$ such that $C$ is invariant under $\pi^p$, where $C$ is the connected component of $G\setminus(X_{args}\cup L_{args})$ that contains $s$, and let $\mathcal{G}_{move} = \FTildePWithZ{t}{x}{l}{k}{z}{p}{M_X}{M_L}{M_K}{M_Z}\setminus\mathcal{G}_{inv}$. To prove this recurrence, we will first describe a map from $\mathcal{G}_{inv}$ to the set of tuples $(G_1,G_2,k',x',l',M,M_X',M_L',\hat C,\hat X',\hat L')$ with the properties given below, and then we will describe a map from $\mathcal{G}_{move}$ to the set of tuples $(G_1,G',x',l',C,M_X',M_L',\hat X',\hat L')$ given below. This will imply that $\tilde f_p$ is at most the summation above. Next, to show that $\tilde f_p$ is at least the summation above, we will describe a map from the set of tuples $(G_1,G_2,k',x',l',M,M_X',M_L',\hat C,\hat X',\hat L')$ to $\mathcal{G}_{inv}$, as well as a map from the set of tuples $(G_1,G',x',l',C,M_X',M_L',\hat X',\hat L')$ to $\mathcal{G}_{move}$. Strictly speaking, to prove these two bounds ($\le$ and $\ge$), it remains to show that all of these maps are injective. See \cref{remark:injective}.

Let $args$ be the arguments of $\tilde f_p$. To see that $\tilde f_p$ is at most the summation above, suppose $G\in\FTildePWithZ{t}{x}{l}{k}{z}{p}{M_X}{M_L}{M_K}{M_Z}$. Let $C$ be the set of vertices in the component of $G\setminus(X_{args}\cup L_{args})$ that contains $s$. Let $X' = N(C)\cap X_{args}$, $L' = N(C)\cap L_{args}$, $M_X' = X'\cap M_X$, and $M_L' = L'\cap M_L$. There are two possible cases: either $C$ is invariant under $\pi^p$, or all of $C$ is mapped to some other component of $G\setminus(X_{args}\cup L_{args})$ by $\pi^p$. If we are in the first case, then let $D$ be the set of vertices of all other components of $G\setminus(X_{args}\cup L_{args})$. Let $G_1 = G[X'\cup L'\cup C]$, $G_2 = G[X_{args}\cup L_{args}\cup D]$, $k' = |C|$, $x' = |X'|$, $l' = |L'|$, $M = C\cap M_K$, $\hat C = C\setminus M$, $\hat X' = X'\setminus M_X'$, and $\hat L' = L'\setminus M_L'$. Note that $|\hat C| = k'-|M|$, $|\hat X'| = x'-|M_X'|$, and $|\hat L'| = l'-|M_L'|$. If $l' = 0$ and $M_X'\subseteq M_Z$, then we know $\hat X'\not\subseteq Z_{args}$ since $G\setminus Z_{args}$ is connected. Since $C$ is invariant under $\pi^p$, this means $N(C)$ is also invariant under $\pi^p$. Therefore, by \cref{lemma:subgraph_auto}, $\pi^p|_{V(G_1)}$ is an automorphism of $G_1$ and $\pi^p|_{V(G_2)}$ is an automorphism of $G_2$. Furthermore, $L'$ is invariant under $\pi^p$ since $L_{args}$ is invariant under $\pi^p$, so $M_X\cup M_L'$ and $M_L\setminus M_L'$ are both invariant under $\pi^p$. Since $|M_X'|\le x'$ and $|M_L'|\le l'$, we have $M_X'\in\mathcal{I}_{x'}$ and $M_L'\in\mathcal{J}_{l'}$. We also have $M\in\mathcal{P}_{k'}$ by an argument similar to the proof for $\tilde g$. Now relabel $G_1$ and $G_2$ in a similar way to \cref{lemma:g_tilde_pi}. By the proof of Lemma 3.12 in~\cite{hebertjohnson2023counting}, we have $G_2\in\widetilde F_p(t,x+l',l-l',k-k',z)$ if $l'<l$, and we have $G_2\in\widetilde G_p(t-1,x+l,k-k',z)$ otherwise. In either case, we have $G_1\in\widetilde G_1(t-1,x'+l',k')$.\footnote{See \cref{footnote:technically} in \cref{lemma:g_pi}.} Therefore, we either have $G_2\in\FTildePWithZ{t}{x+l'}{l-l'}{k-k'}{z}{p}{M_X\cup M_L'}{M_L\setminus M_L'}{M_K\setminus M}{M_Z}$ or $G_2\in\GTildeP{t-1}{x+l}{k-k'}{z}{p}{M_X\cup M_L'}{M_K\setminus M}{M_Z}$, and we have $G_1\in\GTildeOne{t-1}{x'+l'}{k'}{p}{M_X'\cup M_L'}{M}$.

If we are in the second case, where $C$ is mapped to some other component of $G\setminus(X_{args}\cup L_{args})$ by $\pi^p$, then we know $(C,M_X',M_L')\in\mathcal{Q}$ by an argument similar to the proof for $\tilde g$. Let $x' = |X'|$, $l' = |L'|$, $\hat X' = X'\setminus M_X'$, and $\hat L' = L'\setminus M_L'$. As before, if $l' = 0$ and $M_X'\subseteq M_Z$, then we know $\hat X'\not\subseteq Z_{args}$ since $G\setminus Z_{args}$ is connected. Let $C_1 = C$, $X_1' = X'$, and $L_1' = L'$. For $2\le i\le p_C$, let $C_i = \pi^p(C_{i-1})$, $X_i' = \pi^p(X_{i-1}')$, and $L_i' = \pi^p(L_{i-1}')$. Let $G_i = G[X_i'\cup L_i'\cup C_i]$ for $i\in[p_C]$. For the remainder of the graph, let $D = V_{args}\setminus(X_{args}\cup L_{args}\cup C_1\cup\cdots\cup C_{p_C})$, and let $G' = G[X_{args}\cup L_{args}\cup D]$. Now relabel $G_1$ and $G'$ in a similar way to \cref{lemma:g_tilde_pi}. We can see that $\pi^{p\cdot p_C}|_{V(G_1)}$ is an automorphism of $G_1$ and $\pi^p|_{V(G')}$ is an automorphism of $G'$ by an argument similar to the proof for $\tilde g$. By an argument similar to Lemma 3.12 in~\cite{hebertjohnson2023counting}, we have $G'\in\widetilde F_p(t,x+l',l-l',k-p_C|C|,z)$ if $l'<l$, and we have $G'\in\widetilde G_p(t-1,x+l,k-p_C|C|,z)$ otherwise. In either case, we have $G_1\in\widetilde G_1(t-1,x'+l',|C|)$.\footnote{See \cref{footnote:technically} in \cref{lemma:g_pi}.} Therefore, we have either $G'\in\FTildePWithZ{t}{x+l'}{l-l'}{k-p_C|C|}{z}{p}{M_X\cup M_L'}{M_L\setminus M_L'}{M_K\setminus C^{\pi}}{M_Z}$ or $G'\in\GTildeP{t-1}{x+l}{k-p_C|C|}{z}{p}{M_X\cup M_L'}{M_K\setminus C^{\pi}}{M_Z}$, and we have $G_1\in\GTildeOne{t-1}{x'+l'}{|C|}{p\cdot p_C}{M_X'\cup M_L'}{C}$, so $\tilde f_p$ is at most the summation above.

We now show that $\tilde f_p$ is bounded below by this summation. For the first case, suppose we are given $1\le k'\le k$, $0\le x'\le x$, $0\le l'\le l$ such that $0<x'+l'<x+l$, a set $M\in\mathcal{P}_{k'}$, a set $M_X'\in\mathcal{I}_{x'}$, a set $M_L'\in\mathcal{J}_{l'}$, a graph $G_1\in\GTildeOne{t-1}{x'+l'}{k'}{p}{M_X'\cup M_L'}{M}$, a graph $G_2$ such that $G_2\in\FTildePWithZ{t}{x+l'}{l-l'}{k-k'}{z}{p}{M_X\cup M_L'}{M_L\setminus M_L'}{M_K\setminus M}{M_Z}$ if $l'<l$ and $G_2\in\GTildeP{t-1}{x+l}{k-k'}{z}{p}{M_X\cup M_L'}{M_K\setminus M}{M_Z}$ otherwise, a subset $\hat C\subseteq V_{args}\setminus(X_{args}\cup L_{args}\cup M_K)$ of size $k'-|M|$ that contains $s$ if $s\notin M_K$, a subset $\hat X'\subseteq X_{args}\setminus M_X$ of size $x'-|M_X'|$ such that $M'\cup\hat X'\not\subseteq Z_{args}$, and a subset $\hat L'\subseteq L_{args}\setminus M_L$ of size $l'-|M_L'|$. We construct a graph $G$ as follows: Relabel $G_1$ and $G_2$ in a similar way to \cref{lemma:g_tilde_pi}, and then glue $G_1$ and $G_2$ together at $X'\cup L'$ to obtain $G$. By \cref{lemma:supergraph_auto}, $\pi^p|_{V(G)}$ is an automorphism of $G$. By the proof of Lemma 3.12 in~\cite{hebertjohnson2023counting}, we have $G\in\widetilde F_p(t,x,l,k,z)$.\footnote{See \cref{footnote:technically} in \cref{lemma:g_pi}.} Therefore, we have $G\in\FTildePWithZ{t}{x}{l}{k}{z}{p}{M_X}{M_L}{M_K}{M_Z}$.

For the second case, suppose we are given $0\le x'\le x$, $0\le l'\le l$ such that $0<x'+l'<x+l$, a triple $(C,M_X',M_L')\in\mathcal{Q}$, a graph $G_1\in\GTildeOne{t-1}{x'+l'}{|C|}{p\cdot p_C}{M_X'\cup M_L'}{C}$, a graph $G'$ such that 
$G'\in\FTildePWithZ{t}{x+l'}{l-l'}{k-p_C|C|}{z}{p}{M_X\cup M_L'}{M_L\setminus M_L'}{M_K\setminus C^{\pi}}{M_Z}$ if $l'<l$ and $G'\in\GTildeP{t-1}{x+l}{k-p_C|C|}{z}{p}{M_X\cup M_L'}{M_K\setminus C^{\pi}}{M_Z}$ otherwise, a subset $\hat X'\subseteq X_{args}\setminus M_X$ of size $x'-|M_X'|$ such that $M'\cup\hat X'\not\subseteq Z_{args}$, and a subset $\hat L'\subseteq L_{args}\setminus M_L$ of size $l'-|M_L'|$. We construct a graph $G$ as follows: Let $G_2,\ldots,G_{p_C}$ all be equal to $G_1$ initially. Next, relabel $G_1,\ldots,G_{p_C}$ and $G'$ in a similar way to \cref{lemma:g_tilde_pi}. For all $i\in[p_C]$, glue the graph $G_i$ to $G'$ at $\pi^{p(i-1)}(X'\cup L')$, and let $G$ be the resulting graph. We can see that $\pi^p|_{V(G)}$ is an automorphism of $G$ by an argument similar to the proof for $\tilde g$. By an argument similar to Lemma 3.12 in~\cite{hebertjohnson2023counting}, we have $G\in\widetilde F_p(t,x,l,k,z)$.\footnote{See \cref{footnote:technically} in \cref{lemma:g_pi}.} Therefore, we have $G\in\FTildePWithZ{t}{x}{l}{k}{z}{p}{M_X}{M_L}{M_K}{M_Z}$, so $\tilde f_p$ is at least the summation above.
\end{proof}


\subsection{Running time}
\label{sec:running_time_pi}

The running time is dominated by the arithmetic operations needed to compute $\tilde f_p$. In the first part of the recurrence for $\tilde f_p$ (corresponding to when $C$ is invariant under $\pi^p$), there are three summations that sum over at most $n$ possible values (the sums over $k'$, $x'$, and $l'$), and there are three summations that sum over at most $2^\mu$ possible values (the sums over $M$, $M_X'$, and $M_L'$). As for the arguments, there are six arguments whose values are at most $n$ ($t,x,l,k,z,p$), and there are four arguments that are sets of moved points with at most $2^\mu$ possible values ($M_X,M_L,M_K$). The analysis for the second part of the recurrence (when $C$ is not invariant) is similar. Therefore, the overall running time is $O(2^{7\mu}n^9)$.

\subsection{Sampling labeled chordal graphs with a given automorphism}
\label{sec:sampling_with_pi}

The algorithm for $\Call{Sample\_Chordal\_Lab}{n,\pi}$ is given by the following theorem.

\samplingLab*

In \cite{hebertjohnson2023counting}, the authors used the standard sampling-to-counting reduction of~\cite{JVV} to derive an algorithm for sampling labeled chordal graphs from the corresponding counting algorithm. For labeled chordal graphs with a given automorphism, \cref{thm:sampling_pi} follows from \cref{thm:counting_pi} in exactly the same way.

\section{Bound on the number of labeled chordal graphs with a given automorphism}
\label{sec:large_mu_bound}

In this section, we prove a bound on the number of $n$-vertex labeled chordal graphs that have a given automorphism $\pi$, in terms of $n$ and the number of moved points of $\pi$ (\cref{theorem:overall_bound}). To do this, we begin by bounding the number of graphs of this form that have a given perfect elimination ordering (PEO) and a given maximum clique size. In fact, we only need to consider PEOs that delete some maximum clique of $G$ as late as possible. The following observation shows that every chordal graph has such a PEO.

\begin{observation}
\label{obs:const_omega}
For every $n$-vertex chordal graph $G$ with maximum clique size $\omega$, there exists a PEO $v_1,\ldots,v_n$ of $G$ such that $G\setminus\{v_1,\ldots,v_{n-\omega}\}$ is a clique of size $\omega$.
\end{observation}

\begin{proof}
Let $C$ be a maximum clique in $G$. For each $i\in[n-\omega]$, by \cref{lemma:simplicial} we can always find a vertex $v_i$ that is simplicial in $G\setminus\{v_1,\ldots,v_{i-1}\}$ and does not belong to $C$. After that, the remaining $\omega$ vertices of the graph can be added to the PEO in any order, since all of these are simplicial in $G\setminus\{v_1,\ldots,v_{n-\omega}\}$.
\end{proof}

Suppose $G$ is an $n$-vertex chordal graph with automorphism $\pi$ and maximum clique size $\omega$. Imagine that we delete the vertices of $G$ in order according to a given PEO that has the property from \cref{obs:const_omega}. For our purposes, it makes sense to truncate this PEO once the graph that remains becomes a clique of size $\omega$ because at that point, there is only one possibility for the edges of the remaining graph. A \emph{truncated PEO} of an $n$-vertex chordal graph $G$ with maximum clique size $\omega$ is an ordering $v_1,\ldots,v_{n-\omega}$ of a subset $S\subseteq V(G)$ such that $V(G)\setminus S$ is a clique of size $\omega$ and such that $v_1,\ldots,v_{n-\omega}$ can be extended to a perfect elimination ordering $v_1,\ldots,v_n$ of $G$. By \cref{obs:const_omega}, every chordal graph has a truncated PEO.

For most of the lemmas in this section, we will count graphs that have the truncated PEO $1,2,\ldots,n-\omega$. For $n\in\N$, $1\le\omega\le n$, and $\pi\in S_n$, let $\alpha_{\pi}(n,\omega)$ denote the number of labeled chordal graphs $G$ with vertex set $[n]$ and maximum clique size exactly $\omega$ such that $\pi$ is an automorphism of $G$ and such that $1,2,\ldots,n-\omega$ is a truncated PEO of $G$. Let $\mathcal{A}_{\pi}(n,\omega)$ denote the set of such graphs.

We only need to bound the number of graphs with this particular truncated PEO because the number of graphs is the same regardless of which truncated PEO we choose. More formally, suppose $v_1,v_2,\ldots,v_{n-\omega}$ is an ordering of a subset of $[n]$ of size $n-\omega$. For $n$, $\omega$, and $\pi$ as above, let $\beta_{\pi}(n,\omega)$ denote the number of labeled chordal graphs $G$ with vertex set $[n]$ and maximum clique size exactly $\omega$ such that $\pi$ is an automorphism of $G$ and such that $v_1,v_2,\ldots,v_{n-\omega}$ is a truncated PEO of $G$. Note that $\alpha_{\pi}(n,\omega)$ and $\beta_{\pi}(n,\omega)$ have different truncated PEOs. The following observation shows that this makes no difference.

\begin{observation}
\label{obs:fixed_PEO}
Let $n\in\N$, $1\le\omega\le n$, $\pi\in S_n$. We have $\alpha_{\pi}(n,\omega) = \beta_{\pi}(n,\omega)$.
\end{observation}

\begin{proof}
Let $v_1,\ldots,v_n$ be an ordering of $[n]$ that begins with $v_1,\ldots,v_{n-\omega}$, and let $\psi:[n]\to[n]$ be the bijection defined by $\psi(i) = v_i$ for all $i\in[n]$. We have the following bijection between $\alpha_{\pi}(n,\omega)$ and $\beta_{\pi}(n,\omega)$: given a graph $G\in\alpha_{\pi}(n,\omega)$, relabel the vertices of $G$ by applying $\psi$ to the label of each vertex.
\end{proof}

Our goal is now to prove a bound on $\alpha_{\pi}(n,\omega)$. It will be useful to consider two different cases having to do with which points are moved by $\pi$.

\subparagraph*{Two cases for the moved points.} Suppose we are given $n\in\N$, $1\le\omega\le n$, and $\pi\in S_n$. Let $\hat C = \{n-\omega+1,n-\omega+2,\ldots,n\}$. This is the clique of size $\omega$ that would remain after deleting the truncated PEO $1,2,\ldots,n-\omega$ from a graph $G\in\mathcal{A}_{\pi}(n,\omega)$. Let $\mu = |M_{\pi}|$. The permutation $\pi$ falls into one of the following cases:
\begin{enumerate}
    \item $|M_{\pi}\cap\hat C|\le\frac{9\mu}{10}$, i.e., at most $\frac{9}{10}$ of the moved points of $\pi$ belong to $\hat C$
    \item $|M_{\pi}\cap\hat C|>\frac{9\mu}{10}$.
\end{enumerate}
We prove bounds for these two cases in \cref{lemma:retrospective,lemma:moved_in_clique}. However, \cref{lemma:retrospective} assumes $\omega\ge\frac{9n}{20}$ and $\mu\le\frac{4n}{5}$, and \cref{lemma:moved_in_clique} assumes $\omega\le n-\frac{\mu}{5}$. We begin by dealing with the corresponding edge cases (when $\omega$ is small or close to $n$, or $\mu$ is close to $n$) in \cref{lemma:small_omega,lemma:large_omega,lemma:large_omega_2,lemma:large_mu} since these cases are relatively simple and use ideas that will also be useful later on.

\begin{lemma}
\label{lemma:basic_bound}
Let $n\in\N$, $1\le\omega\le n$. The number of labeled chordal graphs $G$ with vertex set $[n]$ and maximum clique size exactly $\omega$ such that $1,2,\ldots,n-\omega$ is a truncated PEO of $G$ is at most $$n^n 2^{\omega(n-\omega)}.$$
\end{lemma}

\begin{proof}
For $0\le i\le n-\omega$, let $\mathcal{A}_i$ denote the set of labeled chordal graphs $G$ with vertex set $\{n-\omega-i+1,\ldots,n\}$ and maximum clique size exactly $\omega$ such that $n-\omega-i+1,\ldots,n-\omega$ is a truncated PEO of $G$. We prove $|\mathcal{A}_i|\le n^i 2^{\omega i}$ by induction on $i$, for all $i\in\{0,1,\ldots, n-\omega\}$. When $i = 0$, there is just one graph in $\mathcal{A}_0$, namely a clique of size $\omega$. Now let $i\in[n-\omega]$ and suppose the statement is true for $i-1$. If we delete the vertex with label $n-\omega-i+1$ from a graph $G\in\mathcal{A}_i$, then the resulting graph (say $G'$) belongs to $\mathcal{A}_{i-1}$. We claim that there are at most $n2^\omega$ possibilities for the neighborhood of this vertex in $G$. The neighborhood must be a clique since this vertex is simplicial in $G$, and thus it is contained in some maximal clique of $G'$. By \cref{lemma:maximal_cliques}, $G'$ has at most $n$ maximal cliques. To choose this neighborhood, we can first choose a maximal clique of $G'$ (which has size at most $\omega$) and then choose a subset of it. Therefore, there are at most $n2^\omega$ possibilities. Since $G'\in\mathcal{A}_{i-1}$, the number of possible graphs in $\mathcal{A}_i$ is at most $n2^\omega|\mathcal{A}_{i-1}|\le n2^\omega n^{i-1}2^{\omega(i-1)}\le n^i 2^{\omega i}$. This gives us the desired bound when $i = n-\omega$.
\end{proof}

We can use \cref{lemma:basic_bound} to bound the number of chordal graphs with small or large values of $\omega$.

\begin{lemma}
\label{lemma:small_omega}
Let $n\in\N$, $1\le\omega<\frac{9n}{20}$. The number of labeled chordal graphs $G$ with vertex set $[n]$ and maximum clique size exactly $\omega$ such that $1,2,\ldots,n-\omega$ is a truncated PEO of $G$ is at most $$n^n 2^{n^2/4-n^2/400}.$$
\end{lemma}

\begin{proof}
By \cref{lemma:basic_bound}, the number of such graphs is at most $$n^n 2^{\omega(n-\omega)} = n^n 2^{(n/2+\Delta)(n/2-\Delta)} = n^n 2^{(n/2)^2-\Delta^2},$$ where $\Delta = \frac{n}{2}-\omega$. We know $\Delta^2>\frac{n^2}{400}$ since $\omega<\frac{9n}{20}$, so this proves the bound.
\end{proof}

\begin{lemma}
\label{lemma:large_omega}
Let $n\in\N$, $\frac{3n}{5}<\omega\le n$. The number of labeled chordal graphs $G$ with vertex set $[n]$ and maximum clique size exactly $\omega$ such that $1,2,\ldots,n-\omega$ is a truncated PEO of $G$ is at most $$n^n 2^{n^2/4-n^2/100}.$$
\end{lemma}

\begin{proof}
Similarly to \cref{lemma:small_omega}, the number of such graphs is at most $$n^n 2^{\omega(n-\omega)} = n^n 2^{(n/2)^2-\Delta^2},$$ where $\Delta = \omega-\frac{n}{2}$ this time. We know $\Delta^2>\frac{n^2}{100}$ since $\omega>\frac{3n}{5}$, so this proves the bound.
\end{proof}

\begin{lemma}
\label{lemma:large_omega_2}
Let $n\in\N$, $\pi\in S_n$, and suppose $n-\frac{\mu}{5}<\omega\le n$, where $\mu = |M_{\pi}|$. The number of labeled chordal graphs $G$ with vertex set $[n]$ and maximum clique size exactly $\omega$ such that $1,2,\ldots,n-\omega$ is a truncated PEO of $G$ is at most $$n^n 2^{n^2/4-n^2/100}.$$
\end{lemma}

\begin{proof}
This follows immediately from \cref{lemma:large_omega} since $\mu\le n$, and thus $\omega > \frac{4n}{5}$.
\end{proof}

Next, we will deal with the case when $\mu>\frac{4n}{5}$ (\cref{lemma:large_mu}). When a graph $\hat G$ has an automorphism $\pi$, it is necessary that every subgraph $G$ of $\hat G$ ``respects'' $\pi$, in the sense that the edges of $G$ obey whichever properties are required by $\pi$. More precisely, this requirement is defined as follows. In \cref{def:respects}, one can think of $G$ as being a subgraph of a graph with vertex set $[n]$ for which $\pi$ is an automorphism.

\begin{definition}
\label{def:respects}
Let $n\in\N$, $\pi\in S_n$, and let $G$ be a labeled graph such that $V(G)\subseteq[n]$. We say $G$ \textbf{respects} $\pi$ if whenever a pair of vertices $a,b\in V(G)$ and a number $0\le p\le n-1$ satisfy $\pi^p(a),\pi^p(b)\in V(G)$, we have the following: $a$ is adjacent to $b$ if and only if $\pi^p(a)$ is adjacent to $\pi^p(b)$.
\end{definition}

For a graph $G$ with vertex set $[n]$, $G$ respects $\pi$ if and only if $\pi$ is an automorphism of $G$. Imagine that we build up a graph $G\in\mathcal{A}_{\pi}(n,\omega)$ by starting with an empty vertex set and then adding each of the vertices $n,n-1,\ldots,1$ to the graph, in that order. It is clear that the current graph must respect $\pi$ at every step of this process.

Next, the following definition will be useful for bounding the number of possible neighborhoods of a vertex.

\begin{definition}
\label{def:retrospective}
Let $n\in\N$, $1\le\omega\le n$, $\pi\in S_n$. Let $1\le a\le n-\omega$, let $V' = \{a+1,\ldots,n\}$, and let $b\in[n]$ be the largest number in the cycle of $\pi$ that contains $a$. For $c\in\N$, we say the element $a$ is \textbf{$\frac{1}{c}$-retrospective} with respect to $\omega$ and $\pi$ if $a\ne b$ and the following holds: for every $M\subseteq V'$ such that $\omega-\frac{\omega}{c}\le|M|\le\omega$, we have $$|\pi^p(M)\cap V'|\ge\frac{\omega}{c},$$ where $p$ is the number such that $\pi^p(a) = b$.
\end{definition}

To see the intuition behind \cref{def:retrospective}, again imagine that we build up a graph $G\in\mathcal{A}_{\pi}(n,\omega)$ by adding the vertices $n,n-1,\ldots,1$ to the graph, in that order. Each time we add a vertex $a$ to the current graph (which has vertex set $V'$), we want to bound the number of possible neighborhoods for this vertex. Suppose vertex $a$ is $\frac{1}{20}$-retrospective. We know vertex $b$ has already been added to the graph since $b>a$. The neighborhood of vertex $a$ is contained in some  maximal clique $M$ of $G[V']$. If $|M|$ is much smaller than $\omega$, then there clearly are few possibilities for this neighborhood. On the other hand, if $|M|$ is close to $\omega$ (at least $\omega-\frac{\omega}{20}$), then the definition of $\frac{1}{20}$-retrospective says that a significant portion of the edges/non-edges from $a$ to vertices in $M$ \emph{can be copied from edges/non-edges incident to $b$ that have already been decided in the past} since $\pi$ is an automorphism. In particular, the relationship between $b$ and every vertex of $\pi^p(M)\cap V'$ has already been decided. Therefore, there are strictly fewer than $n2^\omega$ possibilities for the neighborhood of vertex $a$.

Using this notion, we can prove a bound that is tighter than \cref{lemma:basic_bound} for the case when $\mu>\frac{4n}{5}$. By the above lemmas, we can assume $\omega$ is neither too large nor too small.

Suppose we have been given $n$, $\omega$, $\pi$, and $0\le i\le n-\omega$. From now on, let $V_i = \{n-\omega-i+1,\ldots,n\}$, and let $\mathcal{A}_i$ denote the set of labeled chordal graphs $G$ with vertex set $V_i$ and maximum clique size exactly $\omega$ such that $n-\omega-i+1,\ldots,n-\omega$ is a truncated PEO of $G$ and such that $G$ respects $\pi$.

\begin{lemma}
\label{lemma:large_mu}
Let $n\in\N$, $\frac{9n}{20}\le\omega\le\frac{3n}{5}$, $\pi\in S_n$. Let $\mu = |M_{\pi}|$, and suppose $\mu>\frac{4n}{5}$. We have $$\alpha_{\pi}(n,\omega)\le n^n 2^{\omega(n-\omega)}2^{-\omega n/300+\omega/30}.$$
\end{lemma}

\begin{proof}
There are at least $\frac{2n}{5}$ vertices in the truncated PEO $1\ldots,n-\omega$ since $\omega\le\frac{3n}{5}$. Let $B$ be the set of vertices $j$ from this truncated PEO such that $1\le j\le\frac{2n}{5}$, $j\in M_{\pi}$, and $j$ is not the largest number in the cycle of $\pi$ that contains $j$. We claim $|B|\ge\frac{n}{10}-1$. Indeed, since there are only $\lceil\frac{3n}{5}\rceil$ vertices outside of the set $\{1,2,\ldots,\lfloor 2n/5\rfloor\}$ and we have $\mu>\frac{4n}{5}$, that means at least $\frac{4n}{5}-\lceil\frac{3n}{5}\rceil\ge\frac{n}{5}-1$ of the vertices in the set $\{1,2,\ldots,\lfloor 2n/5\rfloor\}$ are moved by $\pi$. At least half of those that are moved are not the largest number in their cycle of $\pi$, so this implies $|B|\ge\frac{n}{10}-1$.

We proceed by a similar argument to the proof of \cref{lemma:basic_bound}, except we now just want to count graphs for which $\pi$ is an automorphism. We prove the following by induction on $i$, for all $i\in\{0,1,\ldots, n-\omega\}$: $|\mathcal{A}_i|\le n^i 2^{\omega i}2^{-\omega k/30}$, where $k = |B\cap\{n-\omega-i+1,\ldots,n\}|$, i.e., $k$ is the number of vertices in $B$ that belong to the vertex set so far. Since $|B|\ge\frac{n}{10}-1$, this statement with $i = n-\omega$ will give us the desired bound.

When $i = 0$, there is just one graph in $\mathcal{A}_0$, namely a clique of size $\omega$. Now let $i\in[n-\omega]$ and suppose the statement is true for $i-1$. Let $a = n-\omega-i+1$. If we delete the vertex with label $a$ from a graph $G\in\mathcal{A}_i$, then the resulting graph (say $G'$) belongs to $\mathcal{A}_{i-1}$. If $a\notin B$, then there are at most $n2^\omega$ possibilities for the neighborhood of this vertex, as we saw in the proof of \cref{lemma:basic_bound}. Now suppose $a\in B$. In this case, we can prove a tighter bound by showing that this vertex is $\frac{1}{30}$-retrospective. Let $M$ be a subset of $V_{i-1}$ such that $\frac{29\omega}{30}\le|M|\le\omega$. Let $p$ be the number such that $\pi^p(a) = b$, where $b$ is the largest number in the cycle of $\pi$ that contains $a$. We know $|\pi^p(M)| = |M|\ge\frac{29\omega}{30}\ge\frac{87n}{200}$, and there are at most $\lfloor\frac{2n}{5}\rfloor$ vertices in $[n]\setminus V_{i-1}$, so there must be at least $\frac{87n}{200}-\lfloor\frac{2n}{5}\rfloor>\frac{n}{30}\ge\frac{\omega}{30}$ vertices of $\pi^p(M)$ in $V_{i-1}$. Hence vertex $a$ is $\frac{1}{30}$-retrospective.

The neighborhood of vertex $a$ is contained in some maximal clique $M$ of $G'$. If $|M|<\frac{29\omega}{30}$, then there are at most $2^\omega 2^{-\omega/30}$ possible subsets of $M$. If $|M|\ge\frac{29\omega}{30}$, then we know the following, since $G$ respects $\pi$: for every $c\in\pi^p(M)\cap V(G')$, $a$ is adjacent to $\pi^{-p}(c)\in M$ if and only if $b$ is adjacent to $c$ in $G'$. The rest of the neighborhood of $a$ could be an arbitrary subset of $\{d\in M:\pi^p(d)\notin V(G')\}$, but this set has size at most $\frac{29\omega}{30}$ since $a$ is $\frac{1}{30}$-retrospective, so there are at most $2^\omega 2^{-\omega/30}$ subsets of $M$ that could be the neighborhood of $a$. Overall, since $G'$ has at most $n$ maximal cliques, there are at most $n2^\omega 2^{-\omega/30}$ possibilities for the neighborhood of $a$.

Therefore, we have $|\mathcal{A}_i|\le n2^\omega|\mathcal{A}_{i-1}|$ if $n-\omega-i+1\notin B$ (in which case $k$ remains the same when going from $i$ to $i+1$), and we have $|\mathcal{A}_i|\le n2^\omega 2^{-\omega/30}|\mathcal{A}_{i-1}|$ otherwise (in which case $k$ increases by 1). This proves the desired bound.
\end{proof}

We are finally ready to prove a bound for case (1), when $|M_{\pi}\cap\hat C|\le\frac{9\mu}{10}$.

\begin{lemma}
\label{lemma:retrospective}
Let $n\in\N$, $\frac{9n}{20}\le\omega\le n$, $\pi\in S_n$. Let $\mu = |M_{\pi}|$, and suppose $\mu\le\frac{4n}{5}$. Also, suppose $|M_{\pi}\cap\hat C|\le\frac{9\mu}{10}$, where $\hat C = \{n-\omega+1,n-\omega+2,\ldots,n\}$. In this case, we have $$\alpha_{\pi}(n,\omega)\le n^n 2^{\omega(n-\omega)}2^{-\omega\mu/400}.$$
\end{lemma}

\begin{proof}
Let $B$ be the set of vertices $j\in[n-\omega]$ such that $j\in M_{\pi}$ and $j$ is not the largest number in the cycle of $\pi$ that contains $j$. At least $\frac{\mu}{10}$ of the vertices in $[n-\omega]$ are moved by $\pi$ since $|M_{\pi}\cap\hat C|\le\frac{9\mu}{10}$, and among those that are moved, at least half of them are not the largest in their cycle of $\pi$. Therefore, $|B|\ge\frac{\mu}{20}$.

We prove the following by induction on $i$, for all $i\in\{0,1,\ldots, n-\omega\}$: $|\mathcal{A}_i|\le n^i 2^{\omega i}2^{-\omega k/20}$, where $k = |B\cap\{n-\omega-i+1,\ldots,n\}|$. Since $|B|\ge\frac{\mu}{20}$, this statement with $i = n-\omega$ will give us the desired bound.

When $i = 0$, we have $|\mathcal{A}_0| = 1$. Now let $i\in[n-\omega]$ and suppose the statement is true for $i-1$. Let $a = n-\omega-i+1$. If we delete the vertex with label $a$ from a graph $G\in\mathcal{A}_i$, then the resulting graph (say $G'$) belongs to $\mathcal{A}_{i-1}$. If $a\notin B$, then there are at most $n2^\omega$ possibilities for the neighborhood of this vertex. Now suppose $a\in B$. In this case, we claim that vertex $a$ is $\frac{1}{20}$-retrospective. Let $M$ be a subset of $V_{i-1}$ such that $\frac{19\omega}{20}\le|M|\le\omega$. Suppose for a contradiction that $|\pi^p(M)\cap V_{i-1}|<\frac{\omega}{20}$, where $p$ is as above. We have $|\pi^p(M)\setminus V_{i-1}|>\frac{18\omega}{20}$ since $|\pi^p(M)|\ge\frac{19\omega}{20}$, so $\pi^p$ has at least $2\cdot\frac{18\omega}{20}$ moved points. This means $\pi$ has at least $\frac{9\omega}{5}\ge\frac{81n}{100}>\frac{4n}{5}$ moved points, since $M_{\pi^p}\subseteq M_{\pi}$. This is a contradiction since we are assuming $\mu\le\frac{4n}{5}$. Hence vertex $a$ is $\frac{1}{20}$-retrospective. By an argument similar to the proof of \cref{lemma:large_mu}, this implies that there are at most $n2^\omega 2^{-\omega/20}$ possibilities for the neighborhood of $a$.

Therefore, we have $|\mathcal{A}_i|\le n2^\omega|\mathcal{A}_{i-1}|$ if $n-\omega-i+1\notin B$ (in which case $k$ remains the same), and we have $|\mathcal{A}_i|\le n2^\omega 2^{-\omega/20}|\mathcal{A}_{i-1}|$ otherwise (in which case $k$ increases by 1). This proves the desired bound.
\end{proof}

Lastly, we prove a bound for case (2), when $|M_{\pi}\cap\hat C|>\frac{9\mu}{10}$.

\begin{lemma}
\label{lemma:moved_in_clique}
Let $n\in\N$, $\pi\in S_n$. Suppose $1\le\omega\le n-\frac{\mu}{5}$, where $\mu = |M_{\pi}|$. Also, suppose $|M_{\pi}\cap\hat C|>\frac{9\mu}{10}$, where $\hat C = \{n-\omega+1,n-\omega+2,\ldots,n\}$. In this case, we have $$\alpha_{\pi}(n,\omega)\le n^n 2^{\omega(n-\omega)}2^{-\mu^2/100+\mu/10}.$$
\end{lemma}

\begin{proof}
Let $\hat D\coloneqq\{v_1,\ldots,v_{n-\omega}\}$ be the set of vertices outside of $\hat C$. There are at least $\frac{\mu}{5}$ vertices in $\hat D$ since $n\ge\omega+\frac{\mu}{5}$, and there are at most $\frac{\mu}{10}$ moved vertices in $\hat D$ since $|M_{\pi}\cap\hat C|>\frac{9\mu}{10}$. Therefore, there are at least $\frac{\mu}{10}$ vertices in $\hat D$ that are fixed points of $\pi$. Out of all of those vertices in $\hat D\setminus M_{\pi}$, let $B$ be the set of the $\lfloor\frac{\mu}{10}\rfloor$ of these with the largest labels. Note that if $G'$ is a graph with vertex set $\{l,l+1,\ldots,n\}$, where $l$ is the vertex in $B$ with the smallest label, then $|V(G')|\le\omega+\frac{\mu}{5}$.

We prove the following by induction on $i$, for all $i\in\{0,1,\ldots, n-\omega\}$: $|\mathcal{A}_i|\le n^i 2^{\omega i}2^{-\mu k/10}$, where $k = |B\cap\{n-\omega-i+1,\ldots,n\}|$. Since $|B|\ge\frac{\mu}{10}-1$, this statement with $i = n-\omega$ will give us the desired bound.

When $i = 0$, we have $|\mathcal{A}_0| = 1$. Now let $i\in[n-\omega]$ and suppose the statement is true for $i-1$. Let $a = n-\omega-i+1$. If we delete the vertex with label $a$ from a graph $G\in\mathcal{A}_i$, then the resulting graph (say $G'$) belongs to $\mathcal{A}_{i-1}$. If $a\notin B$, then there are at most $n2^\omega$ possibilities for the neighborhood of this vertex. Now suppose $a\in B$. The neighborhood of vertex $a$ is contained in some maximal clique $M$ of $G'$. If $|M|<\omega-\frac{\mu}{10}$, then there are at most $2^{\omega-\mu/10}$ possible subsets of $M$.

On the other hand, if $|M|\ge\omega-\frac{\mu}{10}$, then we have $|V(G')|-|M|\le\frac{3\mu}{10}$ since $|V(G')|\le\omega+\frac{\mu}{5}$. We say a vertex $v\in M$ is \emph{singly moved} if $v$ is moved by $\pi$ and belongs to a cycle $C$ of $\pi$ such that $v$ is the only vertex in $C\cap M$. There are at most $\frac{3\mu}{10}$ moved points of $\pi$ in $V(G')\setminus M$ and at most $\frac{\mu}{10}$ in $[n]\setminus V(G')$, so we have at most $\frac{2\mu}{5}$ moved points in $[n]\setminus M$. This means at most $\frac{2\mu}{5}$ of the vertices in $M$ are singly moved, since each of those must belong to a cycle that lies partly outside of $M$. There are at least $\mu-\frac{2\mu}{5} = \frac{3\mu}{5}$ moved vertices in $M$ in total, so at least $\frac{3\mu}{5}-\frac{2\mu}{5} = \frac{\mu}{5}$ of these are not singly moved. For each of these $\frac{\mu}{5}$ vertices, there is at least one other vertex from the same cycle of $\pi$ that can also be found in $M$. Hence it is sufficient to just choose the relationship to $a$ (i.e., edge or non-edge) for at most half of these vertices, since vertices in $M$ that belong to the same cycle of $\pi$ must have the same relationship to $a$. Once those choices have been made, then all of the other edges/non-edges between $a$ and $M$ have also been determined since $G$ respects $\pi$. Overall, since $G'$ has at most $n$ maximal cliques, there are at most $n2^\omega 2^{-\mu/10}$ possibilities for the neighborhood of $a$.

Therefore, we have $|\mathcal{A}_i|\le n2^\omega|\mathcal{A}_{i-1}|$ if $n-\omega-i+1\notin B$ (in which case $k$ remains the same), and we have $|\mathcal{A}_i|\le n2^\omega 2^{-\mu/10}|\mathcal{A}_{i-1}|$ otherwise (in which case $k$ increases by 1). This proves the desired bound.
\end{proof}

\overallBound*

\begin{proof}
Fix a maximum clique size $1\le\omega\le n$. If $\omega<\frac{9n}{20}$, $\omega>\frac{3n}{5}$, or $\omega>n-\frac{\mu}{5}$, then by \cref{lemma:small_omega,lemma:large_omega,lemma:large_omega_2}, we have $\alpha_{\pi}(n,\omega)\le n^n 2^{n^2/4-n^2/400}$. So from now on, we can assume $\frac{9n}{20}\le\omega\le\frac{3n}{5}$ and $\omega\le n-\frac{\mu}{5}$.

If $\mu>\frac{4n}{5}$, then by \cref{lemma:large_mu} we have $\alpha_{\pi}(n,\omega)\le n^n 2^{\omega(n-\omega)}2^{-\omega n/300+\omega/30}$. We know $\omega(n-\omega)\le\frac{n^2}{4}$ for all values of $\omega$. Therefore, the bound in this case is at most $n^n 2^{n^2/4-3n^2/2000+n/30}$ since $\omega\ge\frac{9n}{20}$. On the other hand, if $\mu\le\frac{4n}{5}$, then by \cref{lemma:retrospective,lemma:moved_in_clique} we either have $\alpha_{\pi}(n,\omega)\le n^n 2^{\omega(n-\omega)}2^{-\omega\mu/400}$ or $\alpha_{\pi}(n,\omega)\le n^n 2^{\omega(n-\omega)}2^{-\mu^2/100+\mu/10}$, depending on which vertices are moved by $\pi$. The first of these two is bounded above by $n^n 2^{n^2/4-9\mu n/8000}$ since $\omega\ge\frac{9n}{20}$.

Combining all of these bounds yields $\alpha_{\pi}(n,\omega)\le n^n 2^{n^2/4-f(n,\mu)}$, where $$f(n,\mu) = \min\left\{\frac{n^2}{400},\frac{3n^2}{2000}-\frac{n}{30},\frac{9\mu n}{8000},\frac{\mu^2}{100}-\frac{\mu}{10}\right\}\ge\frac{\mu^2}{900}-\frac{\mu}{10}$$ since $\frac{3n^2}{2000}-\frac{n}{30}\ge \frac{n^2}{900}-\frac{n}{10}$. There are $n$ possible values of $\omega$, and there are at most $n^n$ possible choices for the truncated PEO of an $n$-vertex graph. Therefore, by \cref{obs:fixed_PEO}, the number of labeled chordal graphs with vertex set $[n]$ and automorphism $\pi$ is at most $n^{2n+1}2^{n^2/4-f(\mu)}$, where $f(\mu) = \frac{\mu^2}{900}-\frac{\mu}{10}$.
\end{proof}

\section{Conclusion}

We built upon the algorithm of Wormald for generating random unlabeled graphs to design an algorithm that, given $n$, generates a random unlabeled chordal graph on $n$ vertices in expected polynomial time. This serves as a proof of concept that one can obtain a sampling algorithm for unlabeled graphs from a $\mathsf{GI}$-complete graph class $\mathcal{G}$ using the following two ingredients: (1) an $\mathsf{FPT}$ algorithm for counting labeled graphs in $\mathcal{G}$ with a given automorphism $\pi$ parameterized by the number of moved points of $\pi$ and (2) a bound on the probability that a labeled graph in $\mathcal{G}$ has a given automorphism. A few potential candidates for this are bipartite graphs, strongly chordal graphs, and chordal bipartite graphs, all of which are $\mathsf{GI}$-complete. An additional open problem would be to design a uniform, or approximately uniform, sampling algorithm ––– either for unlabeled chordal graphs or general unlabeled graphs ––– that runs in expected polynomial time even when we condition on the output graph.

\bibliography{references}

\end{document}